\documentclass[11pt, executivepaper]{article}
\usepackage[utf8]{inputenc}
\usepackage[T1]{fontenc}
\usepackage{natbib}
\usepackage{amsmath}
\usepackage{xcolor}
\usepackage{amsfonts}
\usepackage{graphicx}
\usepackage{enumitem}
\usepackage{geometry}
\usepackage{hyperref}
\hypersetup{colorlinks= true, allcolors=blue}
\setcitestyle{aysep={}}
\begin{document}

\title{\textbf{The Dissipative Approach to Quantum Field Theory: Conceptual Foundations and Ontological Implications}}

\author{Andrea Oldofredi\thanks{Contact Information: Universit\'e de Lausanne, Section de Philosophie, 1015 Lausanne, Switzerland. E-mail: Andrea.Oldofredi@unil.ch}\and{Hans Christian \"Ottinger}\thanks{ETH, Polymer Physics, Department of Materials, Leopold-Ruzicka-Weg 4, 8093 Z\"urich, Switzerland, E-mail: hco@mat.ethz.ch}}

\maketitle

\begin{abstract}
Many attempts have been made to provide Quantum Field Theory with conceptually clear and mathematically rigorous foundations; remarkable examples are the Bohmian and the algebraic perspectives respectively. In this essay we introduce the dissipative approach to QFT, a new alternative formulation of the theory explaining the phenomena of particle creation and annihilation starting from nonequilibrium thermodynamics. It is shown that DQFT presents a rigorous mathematical structure, and a clear particle ontology, taking the best from the mentioned perspectives. Finally, after the discussion of its principal implications and consequences, we compare it with the main Bohmian QFTs implementing a particle ontology.
\end{abstract}
\vspace{5mm}

\begin{center}
 \emph{Accepted for publication in the European Journal for Philosophy of Science}
 \end{center}
\clearpage

\tableofcontents
\vspace{5mm}

\section{Introduction}
\label{Intro}

The Standard Model (SM) of particle physics is currently our most accurate answer to the questions concerning the inherent structure of matter. According to this theoretical framework, atoms and molecules composing ordinary matter and anti-matter are constituted by several families of elementary quantum particles divided into (different generations of) fermions and bosons. The latter, moreover, are the carriers of the electromagnetic, the weak and the strong interactions, three of the four fundamental forces in nature; only gravity is not taken into account by the SM. Furthermore, its predictions are tested with an extreme degree of accuracy, making it the most corroborated theory in history of science. Nonetheless, despite these successes, it is well-known that its mathematical structure, i.e.\ the quantum theory of fields (QFT), has been affected by several conceptual conundra and technical problems since its inception. Let us mention just a few of them:
\begin{enumerate}
   \item QFT has been plagued by divergencies from its very beginning, since its equations contain infinite terms that led to unphysical results and predictions (\cite{Duncan:2012aa}, \cite{Teller:1995aa}). The first of such infinite expressions was found by P. Jordan and dates back to the well-known \emph{Dreim\"annerarbeit}, \cite{Born:1926} (for a historical discussion of the divergencies in early QFT see \cite{Schweber:1994}, Chapter 2). Such problems have formally been solved by the introduction of renormalization methods; nonetheless, from an ontological point of view, the issue of finding a natural and convincing cutoff mechanism is still present.
 
\item QFT does not provide a precise ontological picture of reality, since it is not clear what its fundamental entities are. On the one hand, several no-go theorems have been proven to demonstrate the impossibility of a particle ontology in relativistic quantum field theory (\cite{Halvorson2002}, \cite{Hegerfeldt1998b}, \cite{Malament1996}), on the other hand, it is still an open question whether a field ontology is obtainable (\cite{Baker2009}, \cite{Bigaj:2018}). In addition, despite its predictive power, QFT does not provide physical explanations for \emph{individual} processes taking place at the microscopic level---the dynamics of the theory, in fact, is only concerned with formal methods to compute statistical predictions of observable quantities, as in non-relativistic quantum mechanics (QM) (cf.\ \cite{Falkenburg:2007aa}, Chapter 6). 
   
\item In quantum field theory particles lose the status of primary substances (albeit the particle jargon is still used to describe observed phenomena), being defined as excitations of quantum fields. However, the notion of ``quantum field'' raises several metaphysical issues, being it formally described as an \emph{operator valued quantum field}, i.e.\ a space-time region where a particular set of operators is defined. It is worth noting that, as in the case of non-relativistic quantum mechanics, the operators in QFT do not instantiate precise values prior to observations; rather, one may claim that the net of observables defining a quantum field is a set of potentialities---borrowing Heisenberg's famous expression. To provide further support to this claim, we can also say that from the canonical quantization procedure (anti-)commutation relations are defined, and consequently it is possible to impose uncertainty relations also to the field observables. This implies that the non-classical indeterminate nature of the properties of quantum systems is inherited by quantum fields. Thus, we are left with a metaphysically unclear description of matter (cf.\ \cite{Bell:1986aa}).
In addition, we stress that the very concept of field should be regarded as an idealization; according to its definition, it is assumed that the attributes of a given physical system can change their value from point to point in a continuous space, even at infinitesimally small scales. However, in many physical theories the notion of field breaks down at certain scales as for instance in hydrodynamics, where a field theoretic description is not available at the level of micrometers for gases, and nanometers for liquids. Then, one may expect that also the notion of quantum field can be abandoned in more fundamental physical theories than QFT.

\item QFT is plagued by the conclusions of Haag's theorem and the related existence of infinitely many inequivalent representations of the Canonical Commutation Relations (CCR) (cf.\ \cite{Fraser2008}). In 1955 Rudolf Haag proved that there can be no interaction picture in standard relativistic QFT, since the free and interacting fields cannot act in a common Hilbert space, calling into question the mathematical foundations of the theory (\cite{Haag:1955}). In addition, such theorem has been interpreted as a no-go result showing the impossibility for a particle ontology in QFT (\cite{Fraser2006, Fraser2008}). Furthermore, this theory entails the existence of infinitely many inequivalent representations of the CCR: these representations assign different expectation values to the various families of observables, generating a infinitely many physically inequivalent descriptions among which one has to select the proper representation.

\item The persistence of the measurement problem, which is directly inherited from QM as pointed out by \cite{Barrett:2014aa}. Specifically, Barrett states that a correct understanding of the structure of quantum theory and of entangled states in space-like separated regions in the context of relativistic field theories, depends on the solution one gives to the measurement problem. The conclusions he draws are not optimistic: since currently there are no answers to this problem in relativistic regimes, we lack a solid understanding of relativistic entangled states (and of the basic structure of relativistic QM). Even worse, in their present form the three widely accepted solutions---provided by the many worlds interpretation, Bohmian Mechanics and spontaneous collapse theories---cannot be extended to relativistic settings since either violate relativistic conditions, or are too vague to define what a measurement is in relativistic QM.\footnote{For spatial reasons, the treatment of the quantum measurement problem in the context of the dissipative approach to QFT will be explicitly the subject of a future work; thus, in the present essay we will not discuss this issue any further. Let us say very briefly, however, that DQFT addresses the measurement problem through the notion of unraveling which will be discussed in Section~\ref{DQFT}, treating interactions as jumps. In the present essay we will talk briefly about measurements also in Section~\ref{Discussion}. Nonetheless, a full discussion of the quantum measurement problem in DQFT lies behind of the scope of the present paper.}
\end{enumerate}

Many attempts have been made to avoid such issues and to provide conceptually clear and mathematically rigorous quantum field theories. Remarkable examples can be found in the algebraic and axiomatic approaches to QFT, where several models have been proposed to recover QFT from a restricted set of axioms (\cite{Haag:1964}, \cite{Wightman:1964}, \cite{Osterwalder:1973}); another notable perspective is given by Bohmian Mechanics (BM), where a realistic picture of QFT is obtained starting from a sound ontological basis (\cite{Bell:1986aa}, \cite{Durr2005}, \cite{Colin:2007aa}).

In this paper we present a new alternative, effective formulation of QFT based on a proposal contained in \cite{Ottinger:2017}, where a dissipative approach to QFT (DQFT) is discussed in great detail. Learning the lessons of the axiomatic and Bohmian perspectives, this framework is based on a rigorous mathematical formalism and postulates a clear particle ontology, avoiding the classical problems affecting the standard formulation of QFT. Thus, by supplying a description of objects and processes taking place at the quantum field theoretical regime, DQFT aims to provide a realistic understanding of the ontology of the standard model of particle physics.
Furthermore, this novel formulation of QFT, based on theoretic methods of nonequilibrium thermodynamics, takes at face value the language and the experimental evidence of particle physics, providing the standard model with a consistent description in terms of individual particles. Hence, taking seriously into account the physicists' ``particle'' jargon, we substantiate it with a clear corpuscular ontology. More specifically, as we explain in detail in this essay, nonequilibrium thermodynamics is an ideal framework to describe fundamental interactions, since it provides robust and well-defined evolution equations as well as important additional features, such as a fluctuation-dissipation relation characterizing the stochastic nature of the events taking place at QFT length scales. Indeed, given the inherent stochasticity of the quantum theory of fields, one is expected to enter into the realm of irreversible thermodynamics, where entropy, dissipation and decoherence play a fundamental role.\ In addition, the dynamics of DQFT, being based on a dissipation mechanism consisting in an exchange of particles between a given system and its environment, provides new insights not only concerning the corpuscular ontology of quantum field theory, but also about the interpretational problems raised by the above mentioned Haag's theorem. Finally, a new theory of gravity may be treated within the proposed dissipative quantum field theory (cf.\ \cite{Ottinger:2020, Ottinger:2020b}), making this theoretical framework a new candidate for a unificatory account of all fundamental interactions.
\vspace{2mm} 

The essay is structured as follows: in Section~\ref{DQFT} we review the essential mathematical and physical elements of DQFT, whereas in Section~\ref{Discussion} we discuss its main philosophical and physical implications. Section~\ref{Comparison} will be devoted to the analysis of the pros and cons of this perspective with respect to the currently available alternatives to QFT in the context of Bohmian mechanics.\footnote{In what follows the algebraic QFT will not be discussed since it lacks a clear ontology of matter. Furthermore, such approach currently cannot be considered a proper alternative to QFT, since it does not reproduce any model with (realistic) interactions in spacetime (see \cite{Wallace2011}).} Section~\ref{Conclusion} concludes the paper. 

\section{The Dissipative Approach to Quantum Field Theory}
\label{DQFT}

\subsection{Methodological and Metaphysical Guidelines}

The dissipative approach to QFT is built following a set of methodological and metaphysical guidelines which we consider sound requirements to construct a consistent theoretical framework from both a formal and ontological perspective. As already stated, the principal aim of such a proposal is to formulate an alternative, effective quantum theory of fields capable of solving the major problems affecting its standard formulation starting from clear foundations. Hence, let us present the criteria we assume and employ in this essay\footnote{For an extensive discussion of such criteria see \cite{Ottinger:2017}, Chapter 1.}:

\begin{itemize}
\item We consider mathematical consistency and rigor essential requisites of any robust physical theory. On the one hand, mathematical consistency is a virtue useful in order to propose an empirically adequate physical theory, i.e.\ a theoretical framework able to reproduce the statistics of observed experimental findings avoiding computational deficiencies. On the other hand, it ensures that theories do not lead to contradicting results. Referring to this, it will be shown in this section that the mathematical structure of DQFT is consistent and rigorous being based on a set of clear notions and dynamical equations based on nonequilibrium thermodynamics, which guarantee that the formal machinery employed is not affected by the issues characterizing the standard formulation of QFT; it may be said, in fact, that the purpose of thermodynamics is to characterize and formulate robust equations that make mathematical sense, that is, for which the existence and uniqueness of solutions can be proven.

\item A background finite Minkowski space-time is assumed, since physical phenomena treated by QFT are usually represented as events taking place in relativistic space-time settings. This assumption entails several mathematical consequences; for instance, one can retain the inhomogeneous Lorentz transformations, and therefore, Wigner's classifications of particles in terms of mass and spin, considering them as inherent, fundamental properties of elementary particles (cf.\ Section~\ref{Discussion}). It should be underlined, furthermore, that DQFT is not concerned with the inherent nature of space-time: in what follows we remain agnostic towards its ontology, whose treatment will require a deeper theory with respect to QFT. As a consequence, we consider the latter explicitly an effective theory valid only in a specific range of energy-length scales. Thus, it is possible to consider the choice of such background spacetime as a simplifying assumption.\footnote{It is worth noting that an appropriate infinite-size limit is to be assumed at the end of practical calculations.}
\vspace{2mm}

\item Moving to the metaphysical principles, we aim at providing a realistic picture of the objects and processes taking place at QFT scale. More precisely, we will define precisely what are the theoretical entities representing real objects in the world and their dynamical behaviour in physical space, avoiding the metaphysical indeterminacy affecting standard QFT. So far, it is sufficient to state that such an ontology ensures that the dissipative QFT will have a precise commitment towards the existence of a well-defined set of objects, whose reality is independent of any observation and measurement. Hence, we claim, contrary to a widespread view in the philosophical literature, that it is possible to restore a realistic picture of physical processes taking place in space also in the context of QFT. 

\item In order to tame the conceptual and technical problems arising from the different types of infinities occurring in standard QFT by construction, we assume that according to the dissipative approach such infinities are taken to be only potential, not actual. Therefore, in the present theory we keep the number of quantum particles always finite and countable, so that physical states can be described via a Fock space representation. As we will see in the remainder of the paper, this fact will help us to circumvent the metaphysical implications of the infinitely many inequivalent representations of the CCR. In addition, we introduce restrictions preventing the appearance of divergencies: on the one hand, we consider large but finite volumes of space, i.e.\ a finite universe; this fact consequently imposes a characteristic length scale and an infrared regularization. On the other hand, we take into account a dissipative mechanism which is necessary to have ultraviolet regularization. These assumptions are crucial in order to obtain an empirically adequate and well-behaved theory.\footnote{In principle such limits, exactly as those concerning the infiniteness of our background space, must be taken only at the end of actual calculations. Although this fact entails computational disadvantages, it allows to avoid the problematic issues one encounters dealing with actual infinities in QFT.} 

\item Furthermore, we assume that stochasticity naturally emerges in QFT, since there are several random events in such a framework that can neither be mechanically controlled, nor precisely known, as for instance vacuum fluctuations causing electron-positron pairs which spontaneously appear and disappear. The existence of such events and our inability to known and control them should be considered as a natural source of irreversible behaviour. Hence, this fact motivates to propose an inherently stochastic dynamics for DQFT. Moreover, since the latter is based on arguments taken from nonequilibrium thermodynamics, we must underline that in such context random fluctuations are accompanied with dissipation, irreversibility and decoherence. Thus, it seems natural for a QFT based on nonequilibrium thermodynamics to implement a stochastic dynamics, which also is motivated by experimental evidence and the phenomenology of the quantum theory of fields. 
\end{itemize}

To conclude this preliminary illustration of the guiding principles of DQFT, it is worth stressing again that it is explicitly an effective theory, having a definite characteristic scale lying between $10^{-20}$m, which is the scale of super-colliders, and $10^{-35}$m, which is the Planck scale. Consequently, we model the physical influences due to objects and processes at higher energy scales through a heat bath.
Referring to this, we heavily rely on modern renormalization methods---sharing the arguments in favor of them stated in \cite{Wallace2006}---which are essential tools introduced to tame the already mentioned issues concerning infinities and divergencies and keeping the present theory formally well-defined.\footnote{For technical details concerning renormalization methods see \cite{Wilson:1974}, \cite{Duncan:2012aa}, and \cite{Ottinger:2017}. Interesting philosophical discussions are contained in \cite{Butterfield:2015}, \cite{Teller:1995aa} and \cite{Wallace2006}.}
 
\subsection{The Mathematical Arena: Fock Space Representation, Creation and Annihilation Operators and Fields}

Fock space $\mathcal{F}$, a particular kind of complex vector space with inner product, is the mathematical arena in which the dissipative approach to QFT takes place. In this state space a system of independent quantum objects---whose number can vary in time---is represented by the following expression: 
\begin{align}
\label{Fock}
|n_i\rangle=|n_1, n_2, \dots\rangle.
\end{align}
\noindent The states of the form written above represent an orthonormal basis vector in $\mathcal{F}$, where the ket on the r.h.s.\ indicates a vector in which $n_1$ represents the number of objects in the state 1, $n_2$ represents the number of objects in state 2 and so on. It is worth noting that \eqref{Fock} only counts the number of quantum objects present in a certain state, it does not assign any label to them, i.e.\ these objects do not possess an inherent ``thisness'' or ``haecceity'' using Teller's words; alternatively stated, particles of the same species in the same state are absolutely indistinguishable. 

In addition, for bosons each occupation number $n_i$ is a non-negative integer; for fermions it must be 0 or 1 in virtue of Pauli's exclusion principle, which prevents the possibility for different fermions to occupy the same state. The vacuum state $|0\rangle$ denotes a state in which all occupation numbers vanish, or more precisely, a state in which no object is present. For the sake of simplicity, we will speak about bosons and fermions, however, we will properly introduce the fundamental objects of this theory, i.e.\ its ontology, later on.

Exactly as in standard QFT, creation and annihilation operators for bosons and fermions are defined in $\mathcal{F}$.\footnote{The creation operators permit the generation of all basis vectors in $\mathcal{F}$ from the vacuum state---via its recursive application---while the annihilation operators ``annihilate'' it. For detailed discussions on Fock space see \cite{Ottinger:2017}, Chapter 1, \cite{Teller:1995aa}, Chapter 3 and \cite{Duncan:2012aa}, pp.\,47-48.} In the first case, the the creation operator $a^{\dagger}_i$ increases the number of bosons in the state $i$ by one,  
\begin{align}
a^{\dagger}_i|n_1, n_2, \dots\rangle =\sqrt{n_i+1}|n_1, n_2, \dots, n_i +1\rangle,
\end{align}

\noindent conversely, the annihilation operator $a_i$ decreases it by one:

\begin{align}
a_i|n_1, n_2, \dots\rangle =\begin{cases} \sqrt{n_i}|n_1, n_2, \dots, n_i -1\rangle, \ \text{for $n_i>0$}, 
\\ 
\\
0, \ \text{for $n_i=0$.} \end{cases}
\end{align}

\noindent These operators obey the following commutation relations:
\begin{align}
\label{CCR}
[a_v, a^{\dagger}_{v'}]=\delta_{vv'}
\end{align}
\noindent and 
\begin{align}
\label{CCR_2}
[a_v, a_v']=[a^{\dagger}_v, a^{\dagger}_{v'}]=0
\end{align}

\noindent where $[A,B]=AB-BA$ is the commutator of two generic operators $A, B$ in $\mathcal{F}$. Here we will not consider the definition of such operators for fermions, since these are not strictly relevant for the purposes of the present essay.\footnote{For details see \cite{Ottinger:2017}, p.\,53.} 

It is well-known in the mathematical and physical literature that a Fock space can be rigorously constructed starting from a $N$-particle Hilbert space.\footnote{For instance, \cite{Deckert:2019aa} employed such space to define the Dirac sea picture in Bohmian terms. More details are given in Section~4.2.} The main reason for not following this route to define $\mathcal{F}$ in DQFT is metaphysical in essence, since with the symmetrization and the anti-symmetrization of the tensor products the particles do obtain a label, which is more than what we actually need to define our ontology, as stressed a few lines above. The Hilbert space formalism, thus, ``says too much'' about the inherent nature of quantum particles. On the contrary, the way to define the Fock space presented above eliminates particles' labels, providing us information concerning uniquely the particles' numbers.

In this theory, if we consider a configuration of ``particles'' composed by several species, each of them is represented by an appropriate Fock space; the total configuration will be consequently represented by a single product space, obtained combining each specific Fock space of the individual particles' species at hand. Notably, this latter space will have a unique vacuum, corresponding to the state in which there are no particle of any species. The corresponding Fock states describe an ensemble of independent particles of different kinds; however, not all the possible combinations among states are physically meaningful, as for instance superpositions of boson and fermion states, or states with different electric charges. Such limitations are known as superselection rules.

Furthermore, it is worth stressing that creation and annihilation operators do not carry ontological weight \emph{per se}: they are useful formal tools needed (i) for the definition of a variable number of particles in $\mathcal{F}$, and (ii) to represent physical events of particle creation and destruction occurring in spacetime. Nonetheless, what is ontologically primary in DQFT are quantum particles which can be randomly created and annihilated. These operators, then, play an important functional role, i.e.\ to represent mathematically such physical events. As already mentioned, in DQFT the problem of the infinitely many representations of the canonical commutations relations vanishes by construction, since we have a unique representation of such relations keeping finite the number of the degrees of freedom.

Another step to the definition of DQFT is to select momentum eigenstates to represent single-particle states; as a consequence, momentum space is the fundamental representation of physical systems in this framework. More precisely, we will consider a discrete set of momentum states---this is coherent with the idea to have a Fock space with a countable dimension at any time---on a discrete $d$-dimensional lattice:
\begin{align}
\label{lattice}
K^d=\{ \emph{\textbf{k}} = (z_1, \dots, z_d)K_L | z_j \ \text{integer with $|z_j|\leq N_L$ for all $j=1, \dots, d$}\},
\end{align} 

\noindent where $d$ is the finite dimension of our space, $K_L$ is a lattice constant in momentum space, which is small by assumption, and the large integer $N_L$ limits the magnitude of each component of $\emph{\textbf{k}}$ to $N_LK_L$. In the above equation $K_L, N_L$ are truncation parameters which keep the space finite; in addition, the finite number of elements in $K^d$ correspond to the label $i$ of the general construction of Fock spaces.\footnote{For massless particles the momentum state $\emph{\textbf{k}}=0$ has to be excluded since these objects cannot be at rest, moving at the speed of light.}$^,$\footnote{Taking the the limits $N_L\rightarrow \infty$ (infinite number of particles) and $K_L\rightarrow 0$ (infinite volume) momentum space can be densely covered.}

A further consideration about the ontology of DQFT concerns the role of fields, which do not represent physical entities in spacetime according to the present theory, being only mathematical tools introduced for heuristic reasons without a direct physical meaning. More precisely, they are useful quantities to compute collisions and relevant quantities of interest, but they do not represent physical objects in spacetime in addition to the particles.\footnote{Similar considerations about the heuristic and non-ontological role of fields are present in \cite{Durr2005}, Section 6.} Nonetheless, it is formally useful for the exposition of this theory to introduce the following field (self-adjoint) operator:

\begin{align}
\label{field}
\varphi_\emph{\textbf{x}}=\frac{1}{\sqrt{V}}\sum_{\emph{\textbf{k}}\, \in K^d}\frac{1}{\sqrt{2\omega_k}}\left( a^{\dagger}_\emph{\textbf{k}}+a_{-\emph{\textbf{k}}}\right)e^{-i\emph{\textbf{k}}\cdot \emph{\textbf{x}}}
\end{align}

\noindent where $V$ is the volume of our finite space, $\omega_k=\sqrt{m^2+k^2}$ is a weight factor which is the relativistic energy-momentum relation for a particle of mass $m$.\footnote{In this essay we employ the following units $\hbar=c=1$, where $\hbar$ is the reduced Planck's constant and $c$ is the speed of light.} Interestingly, the physical significance of the factor $1/\sqrt{2\omega_k}$ becomes clear in actual computations of correlation functions of relevant quantities of interest (see Section 2.4). However, it should be stressed that such factors do not permit to interpret the above equation \eqref{field} as a passage from momentum eigenstates to position eigenstates. This fact entails consequences, i.e.\ an indispensable difficulty for DFQT to know where particles are located in space (this problem issue is tamed in the non-relativistic case, where particles have low velocity compared to $c$, $\omega_k$ is substituted with a constant $m$, so that equation \eqref{field} can be interpreted as position eigenstates. We will come back to this issue in Section \ref{Discussion}).

So far we have been silent about what is the ontology of this theory, i.e.\ its fundamental entities, however, on the one hand we have designed the Fock space in a way able to account for individual, discrete, countable objects, whose number can vary in time, on the other hand, we stated that neither creation and annihilation operators, nor fields have ontological status, these are only powerful formal tools appearing in the formal machinery of DQFT.

\subsection{The dynamics of DQFT}

Having defined the state space of our theory, the creation and annihilation operators and fields, let us now discuss two possible ways to describe the dynamics of the dissipative approach to QFT, the first relying on the Schr\"odinger picture, the second on unravelings of a quantum master equation which will be introduced below. Let us start with the former.

In the first place, it is important to underline that the complete dynamics of DQFT represented in the Schr\"odinger picture is composed of two contributions, the reversible and irreversible ones. Considering the reversible contribution, the dynamical evolution of a time-dependent state vector $|\psi_t\rangle$ in Hilbert space which is governed by the well-known unitary Schr\"odinger Equation (SE):

\begin{align}
\label{SE}
\frac{d}{dt}|\psi_t\rangle=-iH|\psi_t\rangle
\end{align}

\noindent where $H$ is the Hamiltonian operator, whose spectrum is assumed to be bounded from below in the context of DQFT. 

In order to describe the full structure of the Hamiltonian, let us take into account the interaction among four colliding particles in a $d$ dimensional space, using the $\varphi^4$ theory\footnote{In the second summand of \eqref{H_coll} we never consider collisions involving more than four particles, since we are exemplifying our theory using a quartic interaction, i.e.\ a local interaction between four fermions at a unique point in space-time.}:

\begin{align}
\label{H_coll}
\begin{split}
&H=\sum_{\emph{\textbf{k}}\, \in K^d}\omega_k a^{\dagger}_\emph{\textbf{k}}a_\emph{\textbf{k}}+\frac{\lambda}{96} \frac{1}{V} \sum_{\emph{\textbf{k}}_1, \emph{\textbf{k}}_2, \emph{\textbf{k}}_3, \emph{\textbf{k}}_4\in K^d} \frac{\delta_{\emph{\textbf{k}}_1+\emph{\textbf{k}}_2+\emph{\textbf{k}}_3+\emph{\textbf{k}}_4, \textbf{0}}}{\sqrt{\omega_{k_1} \omega_{k_2} \omega_{k_3} \omega_{k_4}}}
\\
&\quad \bigl(a_{\emph{\textbf{k}}_1}a_{-\emph{\textbf{k}}_2}a_{-\emph{\textbf{k}}_3}a_{-\emph{\textbf{k}}_4}+4a^{\dagger}_{\emph{\textbf{k}}_1}a_{-\emph{\textbf{k}}_2}a_{-\emph{\textbf{k}}_3}a_{-\emph{\textbf{k}}_4}+6a^{\dagger}_{\emph{\textbf{k}}_1}a^{\dagger}_{\emph{\textbf{k}}_2}a_{-\emph{\textbf{k}}_3}a_{-\emph{\textbf{k}}_4}
\\
&\quad +4a^{\dagger}_{\emph{\textbf{k}}_1}a^{\dagger}_{\emph{\textbf{k}}_2}a^{\dagger}_{\emph{\textbf{k}}_3}a_{-\emph{\textbf{k}}_4}+a^{\dagger}_{\emph{\textbf{k}}_1}a^{\dagger}_{\emph{\textbf{k}}_2}a^{\dagger}_{\emph{\textbf{k}}_3}a^{\dagger}_{\emph{\textbf{k}}_4}\bigr)
\\
&\quad +\frac{\lambda'}{4}\sum_{\emph{\textbf{k}}\, \in K^d}\frac{1}{\omega_k}\left(a_\emph{\textbf{k}} a_{-\emph{\textbf{k}}}+2a^{\dagger}_\emph{\textbf{k}} a_\emph{\textbf{k}} +a^{\dagger}_\emph{\textbf{k}} a^{\dagger}_{-\emph{\textbf{k}}}\right)+\lambda^{''}V.
\end{split}
\end{align}

\noindent In \eqref{H_coll} $\delta$ is Kronecker's $\delta$ and $\lambda, \lambda', \lambda''$ are three free interaction parameters determining the strength of the quartic interaction. More precisely, $\lambda$ should be regarded as the fundamental interaction parameter, whereas $\lambda', \lambda''$ should be considered correction parameters, the former referring to the additional contribution to the square of the mass, and the latter referring to a constant background energy per unit of volume.\footnote{The parameters $\lambda', \lambda''$ become infinite in the limit $N_L\rightarrow \infty$; their actual forms are: 
\begin{align}
\lambda'= \lambda\frac{1}{V}\sum_{\emph{\textbf{k}}\, \in K^d}\frac{1}{4\omega_k},
\\
\lambda''= \frac{1}{2}\lambda\left( \frac{1}{V}\sum_{\emph{\textbf{k}}\, \in K^d}\frac{1}{4\omega_k} \right)^2. 
\end{align}
}

It is important to underline that in \eqref{H_coll} momentum is conserved in collisions; this fact in turn implies the locality of such interactions, which nonetheless does not imply that DQFT has the resources needed to strictly localize particles in space-time, as mentioned above.  

Interestingly, \eqref{H_coll} can change the number of the individual particles by an even amount: 0, meaning that it leaves the number unchanged, $\pm 2$ and $\pm 4$ which means that the particle number can by increased or decreased by 2 and 4 respectively.

The second dynamical contribution of DQFT is inherently stochastic and here is where thermodynamical arguments---more precisely the dissipation mechanism---come properly into the scene. In what follows we describe our physical systems in terms of density matrices $\rho_t$, which can represent a number of different physical states occurring with a certain probability. In this context, density matrices are useful formal tools which enables us to treat ensembles formed by identical and indistinguishable particles, since their statistical properties are completely described in terms of $\rho_t$. It is worth noting that in DQFT density matrices do not represent physical objects in spacetime over and above the ensembles of particles which they describe; in this context they have only a functional role for the particle dynamics. Thus, they should not be compared e.g.\ to the $\psi-$function in Bohm's original pilot-wave theory (cf.\ \cite{Bohm:1952aa}), where the wave function is defined in three-dimensional space, and it is a proper physical field which guides the particles' motion.

Following the usual treatment of dissipative quantum systems, we introduce an inherently stochastic Quantum Master Equation (QME) for the density matrix\footnote{See \cite{Petruccione:2002}, Chapters 3 and 6 for technical details.}, which takes the following form for the above mentioned $\varphi^4$ theory:
\begin{align}
\label{QME_concrete}
\frac{d\rho_t}{dt}=-i[H, \rho_t]-\sum_{\emph{\textbf{k}}\, \in K^d}\beta\gamma_k\int^1_0 e^{-u\beta\omega_k} \left([a_\emph{\textbf{k}}, \rho_t^{1-u}[a^{\dagger}_\emph{\textbf{k}}, \mu_t]\rho^u_t]+[a^{\dagger}_\emph{\textbf{k}}, \rho^u_t[a_\emph{\textbf{k}}, \mu_t] \rho_t^{1-u}]\right)du
\end{align}

\noindent where $a_\emph{\textbf{k}}, a^{\dagger}_\emph{\textbf{k}}$ are the coupling operators which model the interaction between our open system and its environment, which in the present quantum field theory is given by a heat bath of a given temperature $T$, representing the eliminated small-scales/high-energy degrees of freedom which directly influence and interact with our lower energy quantum particles.\footnote{For technical details concerning the general form of the QME see \cite{Ottinger:2011} and \cite{Ottinger:2015}.} Furthermore, the term $e^{-u\beta\omega_k}$ ``produces the proper relative weights for transitions involving the creation and annihilation of free particles'' (\cite{Ottinger:2017}, p.\,65), $\beta=1/k_BT$ represents the inverse temperature, $\gamma_k$ denotes the decay rate, i.e.\ the damping coefficient describing the strength of the dissipation, which is negligible for small $k$ and increases rapidly for large $k$.\footnote{Technical details concerning the justification of the exponential factor $e^{-u\beta\omega_k}$ are given in \cite{Ottinger:2017}, pp.\,66-67.} Here the concrete form of the decay rate\footnote{Other choices of $\gamma_k$ are possible.} is $\gamma_k=\gamma_0+\gamma k^4$: the factor $k^2$ refers to the Laplace operator which causes diffusive smoothing in real space, however the presence of double commutators in \eqref{QME_concrete} suggests the $k^4$ power; the parameter $\gamma_0$ is added since the state with $k=0$ can be subject to dissipation. It is worth noting that the damping of the latter state $k=0$ must be infinitesimally small to be consistent with the results of low energy QFT. Alternatively stated, as the parameter $\gamma$ provides a UV cutoff (note that $\gamma^{1/3}$ defines a length scale), this parameter should be sufficiently small to be in the physically inaccessible range. In the spirit of the renormalization procedure and motivated by the standard procedure in QFT, the precise value of $\gamma$ does not matter here.

In order to characterize the features of the above equation, it is worth noting that the first term on the r.h.s.\ of \eqref{QME_concrete} correspond to the dynamics given by \eqref{SE}. The second term appearing characterizes instead the irreversible process, which is given by the commutators involving the energy operator $\mu_t=H+k_B T\,\mathrm{ln}\rho_t$---the generator of the irreversible dynamics---and the real, non-negative rate factor $e^{-u\beta\omega_k}$. As \cite{Petruccione:2002} p.\,129 underline, the irreversible dynamics is related to entropy production---which is non-negative and vanishes at equilibrium---in the precise sense that the latter is the ``amount of entropy produced per unit of time as a result of irreversible processes''. In addition, as claimed in \cite{Ottinger:2017}, pp.\,62-63: 
\begin{quote}
[t]he multiplicative splitting of $\rho_t$ into the powers $\rho_t^u$ and $\rho_t^{1-u}$, with an integration over $u$, is introduced to guarantee an appropriate interplay with entropy and hence a proper steady state or equilibrium solution. The structure of the irreversible term is determined by general arguments of nonequilibrium thermodynamics or, more formally, by a modular dynamical semigroup.
\end{quote} 

\noindent Consequently, the above mentioned QME implies the convergence to the equilibrium density matrix. Considering this concrete form of QME for a density matrix, it is important to say that temperature is naturally associated with the heat bath that consists of the unresolved, local degrees of freedom of an effective field theory. The ubiquitous loops of collisions involving high-momentum particles and occurring within short periods of time, which are the origin of divergences in QFT, become unresolvable due to the presence of dissipation, which quickly eliminates high-momentum particles and thus provides regularization. A detailed discussion of the resulting unresolvable clouds of individual particles that can be effectively seen through particle detectors can be found in Section~\ref{Discussion}. Referring to this, it is worth noting that in this theory the heat bath constituting the environment is assumed to be in thermal equilibrium, since only the slow large scale degrees of freedom can actually feel the nonequilibrium effects. This fact, in turn, follows from the fundamental assumption of a separation of time scales in nonequilibrium thermodynamics, which entails that the eliminated fine grained degrees of freedom are in equilibrium (cf.\ \cite{Ottinger:2009} for technical details).

Finally, it is important to stress that the QME is one of the most efficient ways to represent the interaction between a quantum system and a heat bath, since an exact treatment of the high energy degrees of freedom would require the solution of a too complex system of coupled equations of motion. In the second place, the evolution of the heat bath's degrees of freedom can be neither known, nor mechanically controlled, thus, one has to simplify the description of such physical situation taking into consideration a restricted set of relevant quantities accounting for this influence. Referring to this, the short-time correlations with the heat bath allow one to neglect memory effects on the dynamics and to define a stochastic Markov process on the state space of the system, given that such times are much smaller than the characteristic time scale of the system's evolution, as clearly stated by \cite{Petruccione:2002}, pp.\,115-122. 

To conclude this section let us briefly underline the crucial role of renormalization group methods in the context of DQFT. As repeatedly stressed, the fast degrees of freedom are eliminated from our theory, these form the environment with which individual particles interact. This scaling is dependent on the friction parameter present in the QME (therefore, also the notion of interacting particle depends on such scaling): increasing the length scale is equivalent to increasing the parameter $\gamma_k$ in \eqref{QME_concrete}, with the consequence of increasing in the entropy production rate. 

Finally, in DQFT the entities that can be subject to detection and observations are clouds of particles emerging from the collisions and interactions of more fundamental and faster degrees of freedom which are instead inaccessible; it should be noted that ``the dissipative coupling to the bath is very weak, except at short length scales. In other words, the dissipative coupling erases the short-scale features very rapidly, whereas it leaves large-scale features basically unaffected'' (\cite{Ottinger:2017}, p.\,29). Furthermore, since there is no a clear cut decoupling among the various high and low-energy processes we assume self-similarity, meaning that although the faster degrees of freedom are eliminated and not directly treated by DQFT, we stipulate that they behave in a similar way with respect to the slower degrees of freedom. Rigorous arguments to justify this claim are contained in \cite{Ottinger:2009}.\footnote{More precisely, we should say that the emergence of irreversible process in nonequilibrium thermodynamics is based on a clear distinction among different levels of descriptions, however, typically one does not have such a clear cut separation. This entails that one can rely on self-similarity instead of a hierarchical view of deeper and deeper layers of reality. Alternatively stated, we are assuming that nonequilibrium thermodynamics works well with both hierarchical and self-similar systems.}

\subsection{Quantities of Interest}

The quantities of interest one may want to compute rest on subjective decisions; however, in this subsection we will provide the most general class of multi-time correlation functions associated to  measurable quantities which connect the general abstract formalism of the theory presented so far with experimental evidence.\footnote{For technical details see \cite{Ottinger:2017}, pp.\,69-72, \cite{Petruccione:2002}, Chapter 3, pp.\,125-128, and \cite{Zoller:2004}.} 

Firstly, it is worth noting that here we deal uniquely with statistical quantities, hence, it is natural to work with density matrices, as previously anticipated. The formal expression of a multi-time correlation function is given as follows: 

\begin{align}
\label{corrfun}
\textrm{tr}\{ \mathcal{N}_nA_n\mathcal{E}_{t_n-t_{n-1}} (\dots \mathcal{N}_2A_2\mathcal{E}_{t_2-t_1} (\mathcal{N}_1A_1\mathcal{E}_{t_1-t_0}(\rho_0)A^{\dagger}_1)A^{\dagger}_2\dots)A^{\dagger}_n\}.
\end{align}

\noindent This formula must be read from the inside to the outside: we start from a density matrix $\rho_0$ at time $t_0$, the evolution super-operator $\mathcal{E}$ is obtained by solving the QME over a definite interval of time $t$, $A_j$ represent linear operators associated with times $t_j$ with $t_0 < t_1 < \dots, < t_n$, finally the normalization factors $\mathcal{N}$ guarantee that after every step the evolution continue with the density matrix. Importantly, the experimental outcomes of a time series of different measurements is contained in the normalization factors.\footnote{Note that the QME \eqref{QME_concrete} is multiplicatively linear but additively nonlinear.}

\subsection{Unraveling of the Quantum Master Equation}

Another possibility to represent the dynamics of the dissipative approach to quantum field theory is based on the notion of \emph{unraveling} of the quantum master equation.\footnote{For details the reader may refer to \cite{Petruccione:2002}, Chapter 6, and \cite{Ottinger:2017}, Section 1.2.8.}  Specifically, instead of formulating the dynamics of DQFT using quantum master equations for density matrices, it is possible to represent it in terms of a stochastic process in the system's state space. 
Thus, the fundamental idea at play is to re-write the dynamics of the presented theory obtaining a time-dependent density matrix $\rho_t$ solving a QME as second moment or expectation $\rho_t=E(|\psi_t\rangle\langle\psi_t|)$, where $|\psi_t\rangle$ is a stochastic process in the relevant Fock space of the open system at hand consisting of periods of continuous Schr\"odinger-type evolution interrupted by random quantum jumps. 
We underline that the unravelings are not unique, and here we explain only the most basic ideas behind unravelings for the simplest case of a non-interacting theory (for more general developments see \cite{Ottinger:2017} and references therein). We first fill in some details on the one-process unravelings considered above and then motivate and develop the idea of two-process unravelings. Here and for all generalizations, we consider unravelings in which the state vector at any time $t$ is a complex multiple of one of the base vectors of ${\cal F}$, where interactions need to be expressed as jumps. This restriction, which can be regarded as a superselection rule, has important consequences: at any time $t$, the system has a well-defined particle content and superpositions do not play any role in our unravelings (cf.\ also \cite{Pashby:2020}). The practical advantages of this restriction for numerical simulations is discussed in Section 3.3.
\vspace{2mm}

\noindent \textbf{One-Process Unraveling}: The main idea can be explained more conveniently by considering the zero-temperature master equation for the non-interacting theory:
\begin{align}
\label{14}
\frac{d}{dt}|\psi_t\rangle=-iH_{\textrm{free}}|\psi_t\rangle-\sum_{k\in K^d}\gamma_k(1-|\psi_t\rangle\langle\psi_t|)a^{\dagger}_ka_k |\psi_t\rangle.
\end{align}

\noindent The above equation contains a dissipative term, and the continuous Schr\"odinger-type evolution \eqref{14} is interrupted by jumps of the form:
\begin{align}
\label{15}
|\psi_t\rangle\rightarrow\frac{a_k|\psi_t\rangle}{\Vert a_k|\psi_t\rangle\Vert}
\end{align}
\noindent occurring with rate $2\langle\psi_t|a^{\dagger}_ka_k|\psi_t\rangle$. For a clear explanation let us consider Figure 1:

\begin{center}
\includegraphics[scale=0.8]{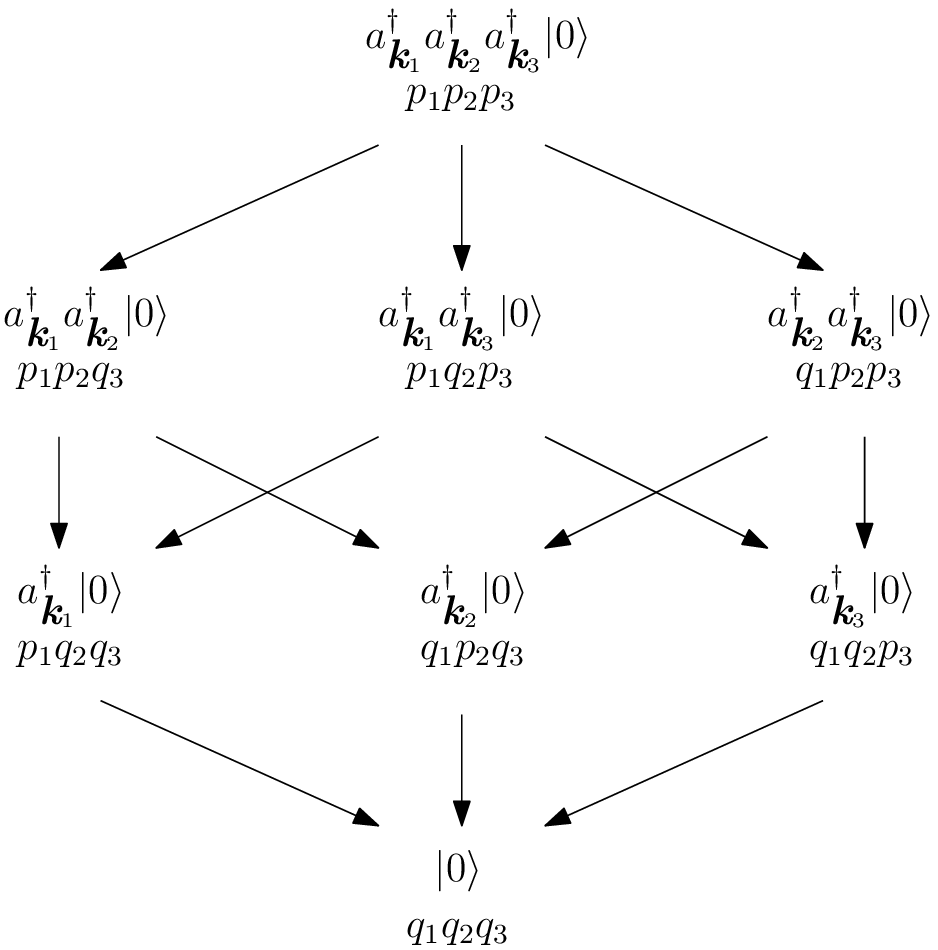}
\begin{quote}

Figure 1: One-process unraveling, where $p_j= e^{-2\gamma_{k_j}t}, q_j=1-p_j$; this picture is taken from \cite{Ottinger:2017}, p.\ 94. NB: this diagram refers to the zero-temperature limit, meaning that the energy goes always down and never up.
\end{quote}
\end{center}

\noindent Here we consider a decay of a three-particle state, particles are then removed until one gets the vacuum $|0\rangle$. Interestingly, at any time $t$ one can calculate the probability to find any state that can be generated by removing one of the particles from the initial Fock space. Looking at Figure 1 we start at the top vertex of the hexagon with three particles, by removing one of them we can obtain three different states represented in the second line; by reiterating the process we obtain three different one-particle states and, eventually, one can reach the vacuum annihilating the last particle. Interestingly, it should be underlined that at any finite time, it is possible to compute the probability to find any state that can be obtained by the removal of a number of particles from the initial Fock state. Since in QFT we have to do with real events of creation and annihilation of quantum objects, we interpret these unraveling as real physical processes in space.
\vspace{2mm}

\noindent \textbf{Two-Process Unravelings:} In the one-process unraveling any change in $|\psi_t\rangle$ affects and modifies in the same way both the bra and ket component of $|\psi_t\rangle\langle\psi_t|$. However, for interacting theories or when we are interested in more general correlation functions than those listed in \eqref{corrfun}, we need to decouple the bra and ket components. In this case, one should use the two-process unraveling, which are based on the following representation of the density matrix of our system $\rho_t=E(|\phi_t\rangle\langle\psi_t|)$, where $|\phi_t\rangle$ and $|\psi_t\rangle$ are two random trajectories in Fock space, i.e.\ two different lists of individual particles, with potentially different jumps.\footnote{If the two vectors $|\phi_t\rangle$ and $|\psi_t\rangle$ are initially equal unit vectors one recovers the one-process unraveling.} For the example of the free theory at zero temperature, the two-process unraveling introduces two simultaneous jumps:

\begin{align*}
|\phi_t\rangle\rightarrow\frac{a_k|\phi_t\rangle \Vert |\phi_t\rangle\Vert}{\Vert a_k|\phi_t\rangle\Vert}
\\
\\
|\psi_t\rangle\rightarrow\frac{a_k|\psi_t\rangle \Vert |\psi_t\rangle\Vert}{\Vert a_k|\psi_t\rangle\Vert}
\end{align*} 

\noindent with rate $2i_k(|\phi_t\rangle, |\psi_t\rangle)\gamma_k$, where $$i_k(|\phi_t\rangle, |\psi_t\rangle)=\frac{\Vert a_k|\phi_t\rangle\Vert \Vert a_k|\psi_t\rangle \Vert}{\Vert |\phi_t\rangle\Vert \Vert |\psi_t\rangle\Vert},$$ and two unitary evolution equations:

\begin{align*}
\frac{d}{dt}|\phi_t\rangle=-iH_{\textrm{free}}|\phi_t\rangle-\sum_{k}\gamma_k\left[a^{\dagger}_ka_k-i_k(|\phi_t\rangle, |\psi_t\rangle)\right]|\phi_t\rangle
\\
\frac{d}{dt}|\psi_t\rangle=-iH_{\textrm{free}}|\psi_t\rangle-\sum_{k}\gamma_k\left[a^{\dagger}_ka_k-i_k(|\phi_t\rangle, |\psi_t\rangle)\right]|\psi_t\rangle.
\end{align*} 

\noindent In this case jumps can take place only if both the two vectors $|\phi_t\rangle$ and $|\psi_t\rangle$ contain a particle with the same momentum $k$.

The two-process unraveling is helpful in the calculation of multi-time correlation functions of a more general type than listed in \eqref{corrfun}, as illustrated in the diagram below:

\begin{center}
\includegraphics[scale=0.8]{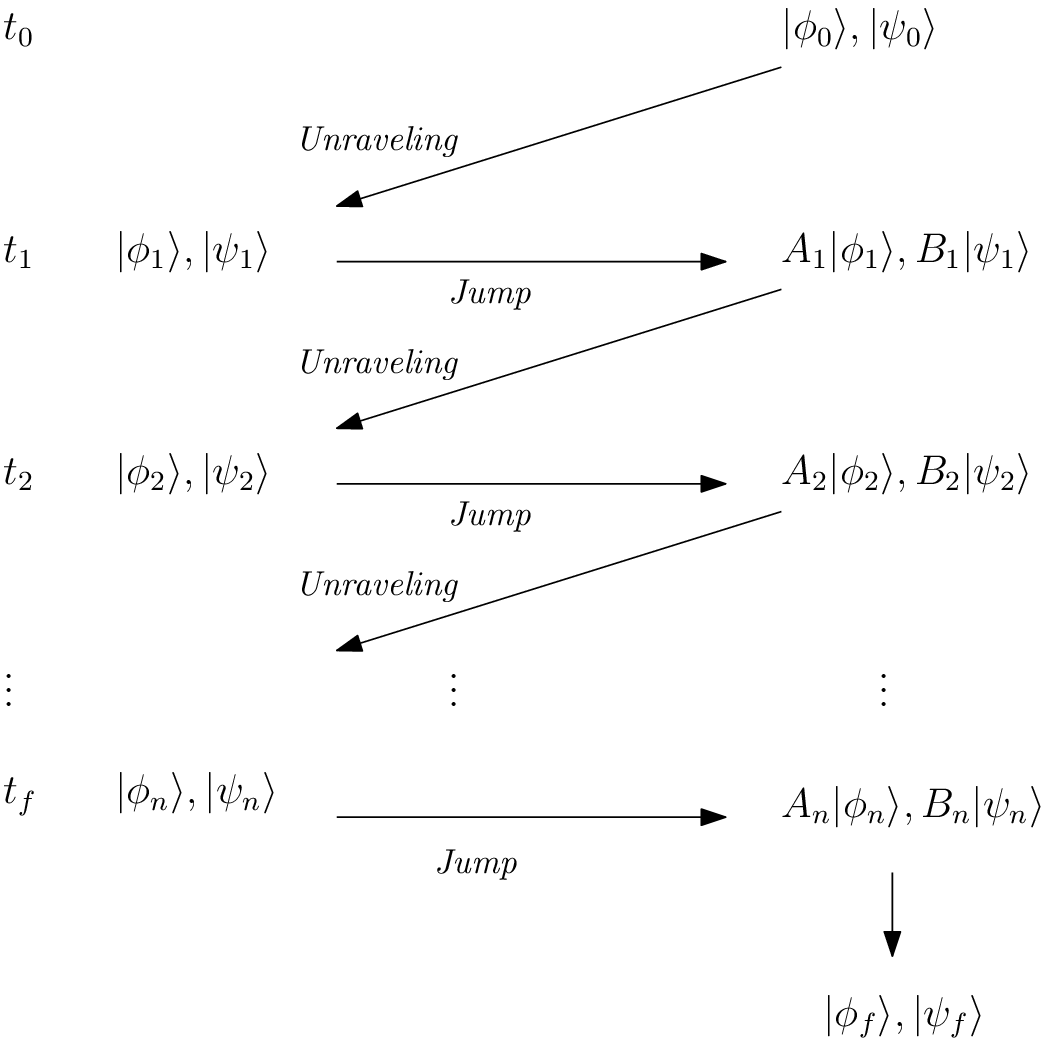}
\begin{center}
Figure 2: Two-process unraveling; picture taken from \cite{Ottinger:2017}, p.\ 97.
\end{center}
\end{center}
\noindent Considering an initial ensemble of states $|\phi_0\rangle$ and $|\psi_0\rangle$ representing the system's density matrix $\rho_0$, they evolve from time $t_0$ to $t_1$ according to the two-process unraveling. The operators $A_i, B_j$ are then introduce via the jumps of $|\phi_j\rangle$ and $|\psi_j\rangle$, at times $t_j$, between these jumps the states and their trajectories in Fock space evolve according to the two-process unraveling. At the final time one gets the final states $|\phi_f\rangle$ and $|\psi_f\rangle$, which allow us to evaluate the multi-time correlation function as follows:
\begin{align*}
tr\bigg\{A_n\mathcal{E}_{t_n-t_{n-1}} \bigg(\dots A_2\mathcal{E}_{t_2-t_1}\bigl(A_1\mathcal{E}_{t_1-t_0}(\rho_0)B^\dagger_1\bigl) B^\dagger_2 \dots \bigg) B^\dagger_n \bigg\}=E[\langle\psi_f|\phi_f\rangle].
\end{align*} 

This discussion can be extended incorporating various forms of the unravelings and concrete examples applied to the $\varphi^4$ theory; a fully detailed picture of these processes are given in \cite{Ottinger:2017}, Section 1.2.8. However, a such technical discussion is beyond the introductory scope of the present essay.

\subsection{Dissipation Mechanism: More than Another UV Regularization Scheme}

To conclude our introduction to DQFT let us play the role of devil's advocate. Considering the dissipation mechanism of DQFT, one might regard it just as another ultraviolet regularization scheme such as, for example, lattice regularization, momentum cut-off, dimensional regularization, or Pauli--Villars regularization. Thus, one would conclude, DQFT would simply retrace the road of standard QFT in order to avoid unwelcome results as those summarized in Section \ref{Intro}. Contrary to this potential objection, in this subsection we are going to explain why dissipation should not be considered another merely formal regularization scheme. In what follows, then, we summarize a number of arguments showing that there is much more to dissipation; some of them will be elaborated in more detail in the following section.

\begin{enumerate}
\item It is worth noting that regularization is deeply related to renormalization, that is, to the elimination of degrees of freedom. Moreover, whenever degrees of freedom are eliminated one should expect entropy and dissipation to play a role, i.e.\ one should expect to enter the realm of irreversible thermodynamics. The occurrence of irreversibility should be considered natural since the infamous divergences in QFT arise from spontaneous particle creation and annihilation, processes that are far beyond our mechanistic control being too fast and too local. This is the motivation which led us to assume that stochasticity naturally emerges in QFT in Section 2.2.

\item Despite the stochastic character of fundamental interactions, they are described via Hamiltonian dynamics (without much critical questioning) which has a pure reversible structure. The equations of irreversible thermodynamics possess a mathematical structure that \emph{generalizes} Hamiltonian dynamics. Nonequilibrium thermodynamics, indeed, not only provides robust evolution equations, but also important additional features, such as a fluctuation-dissipation relation characterizing the thermal fluctuations accompanying a dissipation mechanism at nonzero temperature (see \cite{Ottinger:2020c} and references therein). Hence, the dissipation mechanism seems to be more appropriate to represent fundamental interactions.

\item In the effective field theories of electro-weak and strong interactions, the strength of the dissipation in DQFT is a variable parameter, very much like a lattice spacing or a momentum cutoff, requiring a renormalization treatment. However, unlike these merely computational tools, dynamic dissipative smearing provides a more appealing option for a physical theory at some fundamental scale, namely the Planck scale. Dissipative smearing may be interpreted as the origin of the limit of resolution at the Planck scale and must hence associated with gravity. An alternative theory of gravity that could be treated by means of DQFT has been proposed and elaborated in \cite{Ottinger:2020, Ottinger:2020b}. This higher derivative theory of gravity effectively selects a small subset of solutions from the Yang-Mills theory based on the Lorentz group via constraints. As a result, all fundamental interactions would be unified by DQFT in terms of constrained irreversible dynamic equations under the umbrella of Yang-Mills theories.

\item The dissipation mechanism appearing in the QME (\ref{QME_concrete}) is formulated in terms of the creation and annihilation operators associated with the free Hamiltonian and hence consists of an exchange of particles between the system and its environment, where the exchange of high-energy particles is strongly favored. In our view, this irreversible contribution to dynamics suggests a particle ontology, also in the light of QFT phenomenology. Thus, the formal structure of the theory seems to reflect appropriately the experimental evidence available from particle accelerators.

\item Finally, as we have seen in this section, the formulation of the thermodynamic QME of DQFT relies heavily on the Fock space associated with the creation operators of the momentum eigenstates of the free Hamiltonian, which we interpret as particles. The particle-free state vector $|0\rangle$ of the Fock space may be interpreted as the ground and vacuum state of the free theory. The density matrices obtained from the QME (\ref{QME_concrete}), in which the full Hamiltonian with all interactions is employed in formulating the reversible dynamics, describe the states of the fully interacting theory, including the steady state at a given temperature. In this picture, then, the vacuum states of the free and interacting theories have clearly distinct characters and significance, so that we get new insight into the problems raised by Haag's theorem.
\end{enumerate}

\section{Implications of DQFT}
\label{Discussion}

After introducing the essential mathematical and physical elements of the dissipative approach to QFT, it is possible to describe the basic aspects of this theory as follows. Physical systems are described by a finite but varying number of particles in an appropriate Fock space; referring to this, it is worth noting that in DQFT superpositions of particles are not allowed due to superselection rules already mentioned in the previous section. Against this background, superpositions have only a dynamical origin as a consequence of the conventionally used Schr\"odinger evolution \eqref{SE} (cf.\ \cite{Pashby:2020}). Furthermore, creation and annihilation operators as well as fields are defined. However, they do not possess any ontological status: as already said in the previous section, such mathematical objects have the heuristic, functional role to represent events of particle creation and annihilation, and to simplify actual calculations of collisions and quantities of interest respectively. Thus, although they have formal significance, they do not represent physical objects in spacetime in addition to the particles. Concerning the dynamics of DQFT, the motion in space and time of low energy/slow large scale degrees of freedom is described by a consistent thermodynamic QME for the evolution of a density matrix. Alternatively, it is possible to represent the dynamics of this theory in terms of unravelings of the QME, i.e.\ stochastic processes in the system's state space allowing for quantum jumps. These jumps are spontaneous, random processes in Fock space, and therefore, they are independent of the action of any external measurement or observer. Finally, in this theoretical framework randomness is present not only at the level of the initial conditions, but also in the dynamical evolution of the quantum system, so that DQFT is inherently a stochastic picture of nature. It is our aim now to analyze the main philosophical and physical implications of this theory. 

\subsection{The Particle Ontology of DQFT}

In the first place, it should be noted that DQFT postulates a particle ontology, as anticipated at the end of Section 2.6. This theory is committed, indeed, to the existence of the families of elementary, individual particles accepted by the standard model of particle physics, hence, to the reality of fermions and bosons, which are both considered corpuscular entities.

According to the SM, we have three fermions generations consisting in two quarks and two leptons, the generation of quarks are divided into up/down, charm/strange and top/bottom, while the lepton generation is given by the electron, muon and tau, associated to their neutrino partner. These fermions are massive particles and possess spin-$1/2$. Taking into account bosons, DQFT accepts the existence of gluons, photons, $Z$ and $W^{\pm}$ which are respectively the carriers of the strong, the electromagnetic and weak interactions. These bosons are massless and have spin 1. Finally, the Higgs boson---the only spin-0 particle---is considered real as well. In addition, we postulate that all these particles have inherent properties like mass, charge. Moreover, such particles instantiate also spin, momentum, \emph{etc.}, although their actual values are known solely in measurement situations, given the contextuality of quantum observables (cf.\ \cite{Kochen:1967}).

In the second place, in virtue of what has been stated in Section~2.2 concerning the fundamentality of momentum space---i.e.\ that equation \eqref{field} imposes a stringent formal limitation about particle localization---we should claim that since particles have definite momenta, their spatial location cannot be precisely defined, but at best inferred from measurements. Alternatively stated, albeit quantum particles must be somewhere in space, in the context of DQFT information concerning localization is lost in non-measurement situations. More precisely, in DQFT we can know only that in interactions quantum particles are localized in the very same spatial point, although the positions where such collisions occur is unknown---i.e.\ in this theoretical framework one does not predict where and when a collision is going to happen.


Referring to this, it is worth noting that the most important experimental results from particle accelerators are differential cross sections for certain scattering processes, which are theoretically characterized by suitable correlation functions. Experimental results for cross sections are obtained analyzing many scattering events identified from the particle tracks emerging from points at which high-energy collisions take place. Although these particle tracks may look continuous, they consist of individual points where interactions in the detector take place and one of the emerging particles gets detected.\footnote{It should be pointed out that in principle there might be gaps in particles' trajectories. Nonetheless, if the gaps were (too) large, we would be confused since they would somehow affect the particles' trajectories, and end up in something different with respect to the observed paths in particle accelerators.}
In more detail, taking into account the available experimental evidence, we typically have a high-energy collision followed by many low-energy collisions serving as detection events, so that a high-energy particle can be traced without changing the properties of the particle too much. These individual points where particles collide are the observable events in DQFT, and the only source of information concerning particles' location in spacetime---i.e.\ only in such detection events we can affirm that the colliding particles have a precise spatial localization. It should be noted, furthermore, that these arguments rely only on the interactions being local, and we should not make any stronger assumption about the particle positions when they are not interacting. 
In this manner DQFT is able to explain the experimental evidence which speaks about particles following trajectories in space, although such particles are not strictly localized objects in such a framework. Hence, we \emph{pragmatically} assume that in DQFT particles exist in space-time also between collision events, although information concerning the exact localization of the particles is not available.

We recall that \cite{Bell:1986aa} postulated an ontology of fermion number density at each point of a discrete lattice space. Hence, although he assumed the existence of particles, his theory does not provide a direct information concerning particles' localization, as for instance in standard Bohmian mechanics. However, spatial information could be in principle implicitly inferred knowing the position of the lattice (which may be very fine). Here we provide a strategy similar to Bell's, inferring indirectly that the particles have space-time location also between interactions, although such information is not provided by the theory. DQFT, therefore, implements a particle ontology without generating formal contradictions with respect to the several no-go theorems proving the impossibility of such ontology in relativistic QFT (e.g. the already mentioned results of \cite{Halvorson2002}, \cite{Malament1996}, \cite{Hegerfeldt1998b}), since its fundamental objects are not formally strictly localized in space-time. More precisely, DQFT does not have the resources to describe particles' locations; this is the reason for which such theory nicely conforms to these theorems, which work by showing that if a certain set of conditions is met by a theory, then such framework does not possess any physical quantity represented by self-adjoint operators (or POVMs in a more general case) representing the location of particles. It should be underlined, however, that in the present essay we share the pragmatic line of thought contained in \cite{MacKinnon2008}, where the author scales down the metaphysical significance of such no-go theorems in the light of the experimental evidence of particle physics. In particular, MacKinnon emphasizes that the standard model assumes a particle language which provides a basic ontological commitment towards the existence of countable particles which can be localizable in experiments, even though it is not possible to construct wave functions with compact support in position basis or proper position operators as in non-relativistic quantum mechanics. Such conclusions, thus, are justified by the extraordinary empirical success of the standard model, and are shared by DQFT, giving ontological substance to the physicists' particle jargon. In addition, in the light of QFT phenomenology, another argument favoring a particle ontology can be found from a realistic interpretation of the dissipation mechanism appearing in \eqref{QME_concrete} which, as already stated in Section 2.6, consists in an exchange of actual physical particles between the system and its environment (i.e.\ the heat bath), where the exchange of high-energy particles is strongly favored. This irreversible contribution to dynamics also suggests that a particle ontology is a strong candidate to explain the available experimental evidence speaking about particles in spacetime.

It is then natural to adopt Fock space as the mathematical arena of DQFT, given that an ontology of individual particles can be supported only via a privileged Fock space representation. These particles are individual countable objects, and considering different species of particles, objects of the same species in the same state are identical. Hence, the theory states precisely what are its fundamental objects, i.e.\ individual particles moving in physical space, which can be randomly created and annihilated. In the present essay we tried to avoid the usual jargon of perturbation theory, which introduces the categories of ``free'' and ``interacting'' particles, since it may suggest that there are two ontologically different types of particles, or more precisely, that free particles cannot interact. On the contrary, in DQFT particles are individual objects singled out by the Fock space which do interact and collide. The expression ``clouds of particles'' simply indicates a group of particles interacting with one another at a precise spacetime point, and by the very fact that they interact, we can know they are in the same spatial location.

Another aspect of the theory which is worth mentioning concerns the QME introduced in the previous section, which describes collisions among individual particles and their interaction with a heat bath, incorporating the influence of higher energy degrees of freedom on these individual particles. Such a dissipative interaction with the heat bath leads to diffusive smearing, or a lack of resolution, suggesting the notion of unresolvable clouds of individual particles. As the particles have well-defined momenta and hence cannot be localized in space, the notion of clouds calls for a more detailed explanation. If a particle is involved in a collision leading to high-momentum particles, in view of the rate parameter $\gamma_k = \gamma k^4$ occurring in the QME, any particles with large momenta $\emph{\textbf{k}}$ are removed very quickly. Although Eq.~(\ref{field}) is meant only for heuristic arguments and, in the relativistic case, is not exactly a Fourier transform, it suggests that the dissipative elimination of high-momentum contributions may be interpreted as spatial smearing or a lack of spatial resolution. As we know that collisions among individual particles are strictly local events, we consider them for a better understanding of dissipative smearing. If, for example, a charged particle emits and reabsorbs a photon, that influences the interaction of the charged particle with other particles. Since the total momentum is conserved in the loop of collisions, the modified interaction is still local. If high-momentum particles are produced in such a collision loop, they can quickly be swallowed by the heat bath, so that the loop cannot be closed, momentum is no longer conserved and the modified interaction with other particles appears to be nonlocal. This nonlocal interaction is interpreted in terms of dissipative smearing and clouds of individual particles; thus, the unresolvable clouds of particles can be regarded as the particles of the interacting theory that is regularized by dissipation.

In sum, individual particles in clouds cannot be observed due to the dissipative smearing appearing in the QME, but they constitute the fundamental building blocks of DQFT.

Moreover, it is possible to underline another remarkable ontological difference between standard QFT and DQFT, since in the latter approach fields do not have ontological meaning being exclusively a mathematical, heuristic tool to compactly express the collision rules without any reference to real objects in the world. By construction, then, every ontological issue generated by the notion of quantum field is avoided. According to DQFT the world is composed by corpuscular objects in motion in space-time obeying a stochastic dynamics, giving to QFT the shape of a mechanical theory. Moreover, measurements or external observers do not play any role in this theory. Therefore, DQFT can be properly considered a theoretical framework providing a definite metaphysical picture of the objects and processes taking place at the QFT level. Consequently, DQFT shows that also effective theories can implement a clear ontology.

\subsection{Haag's Theorem and the Role of Renormalization Methods}

Given that the number of individual particles and momenta is always kept finite (although it can vary), DQFT by construction avoids the appearance of inequivalent representations of the CCR. Related to this, such a theory provides a powerful answer to the ontological implications of Haag's theorem. According to this result, there exists no picture for relativistic QFT since free and interacting fields do not share a common Hilbert space. Specifically, there is no interaction picture available. It is worth noting that this theorem has been used to argue against the possibility of a particle ontology for QFT (\cite{Fraser2006, Fraser2008}): since the Fock space representation available for the free fields cannot be extended to the treatment of interacting fields, it follows that there are no Fock space representations for interacting field theories. Nonetheless, such conclusion can be avoided making use of RG methods which restore a finite number of degrees of freedom as stated in \cite{Duncan:2012aa}. Indeed, the purely mathematical problem of representing the free and full Hamiltonians and their respective ground states (with finite ground state energy) is avoided in DQFT in a conventional manner by keeping the Fock space finite and passing to the limit of infinitely many degrees of freedom only after calculating the quantities of physical interest.\ However, a more interesting solution to the \emph{ontological} issues raised by Haag's theorem can be given in the context of DQFT, since the interplay between the free and interacting theories is more profound than in standard QFT, where it is merely associated with the use of perturbation theory or the interaction picture for solving or simplifying equations. For the reversible contribution to the QME (\ref{QME_concrete}), only the full Hamiltonian $H$ of the interacting theory matters, just as in the standard approach. In the irreversible contribution, however, in addition to the full Hamiltonian $H$ (which enters through $\mu_t$), the creation and annihilation operators of individual particles appear because the heat bath acts on such particles rather than \emph{clouds} of particles. Elements of the free and interacting theories appear in very distinct ways in the QME (\ref{QME_concrete}). The ground state of the Fock space may be interpreted as the vacuum state of the free theory, which is simply devoid of individual particles. The full theory is described by the density matrix $\rho_t$ for which the thermodynamic QME leads to a well-defined stationary state depending on temperature. This equilibrium state should be regarded as the vacuum state of the full theory, which is bubbling with quantum and thermal fluctuations (cf.\ \cite{Auyang:1995}, p.\,151). Hence, the irreversible contribution to dynamics eliminates all concerns about the proper interplay between the free and interacting theories and their vacuum states that are usually associated with Haag's theorem. Alternatively stated, the conventional problem of Haag's theorem is related to the fact that the vacuum of the free picture and the interaction picture are defined in two different Hilbert spaces, whereas in the context of DQFT there is just one, single state space: the vacuum state of the free theory is just the empty Fock space, and the vacuum of the interacting theory is the state of thermal equilibrium in the very same Fock space. More precisely, as clearly stated in the previous section, the formulation of the thermodynamic QME \eqref{QME_concrete} is based on the Fock space associated with the creation operators of the momentum eigenstates of the free Hamiltonian, which in this theoretical framework are interpreted as physical particles. Consequently, the particle-free state vector $|0\rangle$ of the Fock space may be interpreted as the ground and vacuum state of the free theory. On the other hand, the density matrices obtained in the QME \eqref{QME_concrete}---in which the full Hamiltonian with all interactions is employed in formulating the reversible dynamics---describe the states of the fully interacting theory, including the steady state at a given temperature. In this picture, then, the vacuum states of the free and interacting theories have clearly distinct characters.
In our view, dissipative regularization is a much deeper answer to the problems associated with such theorem than truncation to finite-dimensional spaces. In short, the distinction between reversible and irreversible contributions to dynamics requires separate ingredients from the interacting and free theories and leads to a clear conceptual difference between the vacuum states of these respective theories.

Taking into account instead the role of renormalization methods in DQFT, it is worth noting that problems deriving from the existence of actual infinities are circumvented via the introduction of (i) dissipation mechanism (which provides ultraviolet cutoff), and (ii) a large but finite volume of space (introducing infrared cutoffs). These facts guarantee the empirical adequacy of the present model, and naturally defines it as an effective theory, whose validity is strictly confined to its characteristic scale mentioned in the previous section. As already stated in section 2.2, DQFT crucially relies on potential infinities related to limiting procedures. Specifically, we consider a finite space, i.e.\ a finite system volume with a finite number of moments states providing a low-energy, infrared cutoff. Moreover, this framework relies on two different mechanisms leading to the ultraviolet cutoff: on the one hand, there is the dissipative coupling with the heat bath, on the other hand, another ultraviolet cutoff to maintain finite the number of momentum states is introduced being useful for intermediate calculations. Interestingly, this latter cutoff becomes irrelevant at the end of practical computations, given that the dissipative coupling will suffice to prove the desired ultraviolet regularization. Therefore, to keep DQFT well-behaved one has to perform two different limits: the limit of infinite volume, leading to a continuum number of momentum states, and the limit of vanishing friction parameter $\gamma$. Remarkably, the volume $V$ should be smaller of the volume of the entire universe and $\gamma^{1/3}$---larger than the Planck length; however, if a particular theory is formulated for such extreme values would perhaps require an adequate treatment of gravity in QFT. This case, however, lies beyond the scope of the present essay. It should be also noted that in this theoretical framework, such limits are motivated from the metaphysical requirement to avoid actual infinities in our physics. Therefore, we take seriously into account renormalization methods considering it not as a mere formal trick to eliminate divergencies, but rather as a systematic procedure to find well-defined theories and perturbation expansion.

Against this background, it should also be emphasized that Lorentz covariance is not immediately manifest from the equation presented in Section \ref{DQFT}. On the contrary, the assumptions of finiteness of space on the one hand, and the dissipative mechanism on the other, would imply a violation of the principles of special relativity. Nonetheless, in view of our metaphysical criterion according to which one should avoid actual infinities in physical theories, Lorentz invariance may be considered an idealization arising taking certain limits, i.e.\ if we assume that universe is finite in space and time, then Lorentz symmetry is only an approximation which for all practical purposes can be considered exact. What is important in DQTF is that, at the final stage of calculations, this symmetry is respected so that it would be empirically not distinguishable with respect to a genuine Lorentz invariant theory.\footnote{This strategy is also followed by the Bohmian QFTs, since these theories make predictions which are statistically equivalent w.r.t.\ those of standard QFT, albeit they are not genuine Lorentz invariant frameworks.} It is important to underline that Lorentz symmetry goes with Minkowski spaces, that is, with infinite homogeneous continuous metric spaces. If something happens at small or large length scales, Lorentz symmetry must become an approximate one. For accelerator experiments in particle physics, the finite size of the universe is not expected to matter and, therefore, should not spoil the practical validity of Lorentz symmetry. Similarly, it is practically irrelevant whether the UV cutoff mechanism is Lorentz covariant or not. For mathematical reasons, it may nevertheless be convenient to have a Lorentz covariant cutoff mechanism---in particular, in the manifestly Lorentz covariant Lagrangian approach. This can, for example, simplify perturbation theory considerably. However, in a Hamiltonian approach like DQFT, where Lorentz invariance is not manifest anyway, a Lorentz invariant UV cutoff seems to be non-essential. When the dissipative mechanism is considered at a more fundamental scale, as for instance in a theory of gravity, then one might want to develop a Lorentz covariant description of dissipation. In a classical (i.e.\ non-quantum) setting, this has been done in Chapter 5 of \cite{Ottinger:2005}. For the thermodynamic QME, this remains to be done in the same spirit.

Finally, let us stress an important metaphysical feature of the present theory related to the arrow of time. DQFT naturally generates a preferred direction of time's arrow at the characteristic scale in which it is defined, since the dissipative mechanism---introducing irreversible behavior---leads to increasing entropy in time. However, it is not our intention to argue that the irreversibility of our macroscopic world is somehow derived from such fundamental irreversibility, since many dissipative mechanisms can emerge at different length scales. Indeed taking into consideration a dissipative equation with entropy production as the fundamental QME of DQFT, we note that it comes with an arrow of time. However, in the spirit of statistical nonequilibrium thermodynamics, it is natural to consider evolution equations resulting from the fundamental QME upon \emph{further coarse graining}. In general, such equations contain two types of dissipative phenomena: the dissipation inherited directly from the fundamental QME, and additional dissipative processes emerging from such a coarse graining. Consequently, even if the QME sets the arrow of time, it does not directly account for \emph{all} dissipative processes in the macroscopic world. For example, the viscosity of water is not inherited from the dissipative properties of the QME, but it rather emerges in the same way as it would arise from the standard model in its reversible form. In sum, the fundamental dissipation provides UV regularization and should not affect any macroscopic properties, but it sets the direction of time's arrow \emph{at relevant scales}.

\subsection{Numerical Simulations}

In conclusion, among the novelties of this approach we stress that it provides new tools and methods for numerical simulations (cf.\ \cite{Ottinger:2017}, Sections~1.2.8.6 and 3.4.3.3). In particular, the idea of unravelings described in Section~2.3 suggests to solve the fundamental QME (\ref{QME_concrete}) by simulating pairs of stochastic trajectories $|\phi_t\rangle$, $|\psi_t\rangle$ in Fock space, from which the density matrix can be obtained as the average $\rho_t=E(|\phi_t\rangle\langle\psi_t|)$. This idea is particularly useful in the low temperature limit where the QME becomes linear (although inhomogeneous, so that a separate equilibration is required). As already mentioned, a stochastic trajectory consist of intervals of deterministic, continuous time evolution interrupted by a sequence of random jumps, whose necessity arises from the irreversible contribution to dynamics (``thermal fluctuations''); notably, whereas the two members of a pair of trajectories share the same continuous evolution, the jumps are correlated but different. Once jumps have been introduced, it is conceptually very natural and practically very efficient to treat also interactions by jumps (``quantum fluctuations''). By doing so, one can find an unraveling in terms of jumps between the natural base vectors (\ref{Fock}) of the underlying Fock space with prefactors evolving deterministically in time. This option is particularly attractive because it suggests that an unraveling consists of correlated pairs of fluctuating lists of particles. Such an intuitive interpretation is useful because the efficiency of simulations depends on proper importance sampling. Most notably, it is important (i) to control the distance between the two trajectories in a pair to sample the relevant contributions to the correlation functions of interest and (ii) to keep the explored part of Fock space from growing exponentially to avoid the famous sign problem of quantum simulations (see \cite{Loh:1990} and \cite{Troyer:2005}). The striking advantages of stochastic simulations based on unravelings compared to the usual Monte Carlo simulations of lattice quantum field theories\footnote{The foundational ideas of lattice QFT are contained in the seminal essay \cite{Wilson:1974aa}, whereas \cite{Duane:1986} and \cite{Gottlieb:1987} employ them in actual simulations.} originate from the possibility of intuitive importance sampling and from the fact that the new simulation proceeds by jumps, which, on average, occur proportional to physical time, whereas time usually is one of four lattice dimensions so that the Monte Carlo iteration time introduces an extra dimension.

\section{A Comparison with Bohmian QFTs}
\label{Comparison}

After introducing the dissipative QFT and discussing its major implications, in this section we compare it with the most developed and best known Bohmian QFTs implementing a primitive ontology of particles\footnote{For an introduction to the various Bohmian QFTs see \cite{Struyve:2010aa}. The notion of primitive ontology is introduced in detail in \cite{Allori:2013ab}; in the present essay we assume that the reader has some familiarity with the primitive ontology programme.}, the Bell-type QFT (BTQFT henceforth) and the Dirac Sea picture (DS). We will argue that DQFT is able to overcome some difficulties plaguing the mentioned theories.

\subsection{DQFT and Bell-type Quantum Field Theory}

The building blocks of the Bell-type QFT (\cite{Durr:2004aa, Durr2005})\footnote{This theory is (i) a generalization of Bell's model for QFT (cf.\ \cite{Bell:1986aa}), and (ii) an extension of BM to QFT; for details on Bohmian mechanics see \cite{Durr:2013aa}.} are particles (with definite positions) in motion in physical space; contrary to Bohmian mechanics, in this theory trajectories randomly begin and end at certain space-time points. More specifically, according to BTQFT creation events correspond to the beginning of a given trajectory, whereas annihilation events correspond to its end. In this framework phenomena of particle creation and annihilation are literally interpreted, hence, such theory postulates a particle ontology where objects can come randomly into existence, and similarly cease to exist. Then, it follows that the particle number is variable. These jumps are specifically introduced to explain the QFT phenomenology, since experimental evidence suggests that there are literal creation and annihilation of particles.

BTQFT describes physical systems by a pair $(Q, \Psi_t)$, where the former element represents an actual configuration of particles with definite positions, the latter is the state vector which belongs to an appropriate Fock space. As already said, the dynamics of the theory introduces stochastic variations in the particles' number to account for creation and annihilation events, which are assumed to be spontaneous, primitive facts of nature, i.e.\ not caused by any physical processes, external observers or forces. In this framework the state vector evolves according to the Schr\"odinger equation \eqref{SE}, where $H$ is the Hamiltonian, which now can be defined conveniently as the sum of the free and the interacting terms, $H=H^{\textrm{free}}+H^{\textrm{int}}$, where the former represents continuous processes, and the latter describes interactions. In BTQFT, between creation and annihilation events, Bohmian particles evolve deterministically according to the guiding law
\begin{align}
\frac{dQ_t}{d_t}=v^{\Psi_t}(Q_t),
\end{align}
\noindent which depends on the free Hamiltonian. The discontinuities in particles' trajectories are represented via jump rates $\sigma(q', q, t)=\sigma^{\Psi_t}(q', q)$, which involve the $H^{\textrm{int}}$ term. These stochastic jumps describe the transitions from a certain configuration of particles $q$ to another configuration $q'$ which has a different particle number.\footnote{For details on $H^{\textrm{int}}$ see \cite{Durr2005}, Section~2.6.} 

Figure (a) below represents the emission of a photon at time $t_1$ (dashed line) from an electron trajectory (solid line), and its subsequent absorption at time $t_2$ by a second electron; such events correspond to creation and annihilation of the photon respectively. Between them it follows a continuous trajectory, exactly as the electrons. The photon emission implies a jump rate $\sigma$ where the starting configuration is composed by two electrons, and the arriving configuration includes also the photon. Similarly, picture (b) represents a creation of an electron-positron pair at time $t_1$ from the trajectory of a photon, which ends when the particle pair is created. 
\vspace{2mm}

\begin{center}
\includegraphics[scale=0.7]{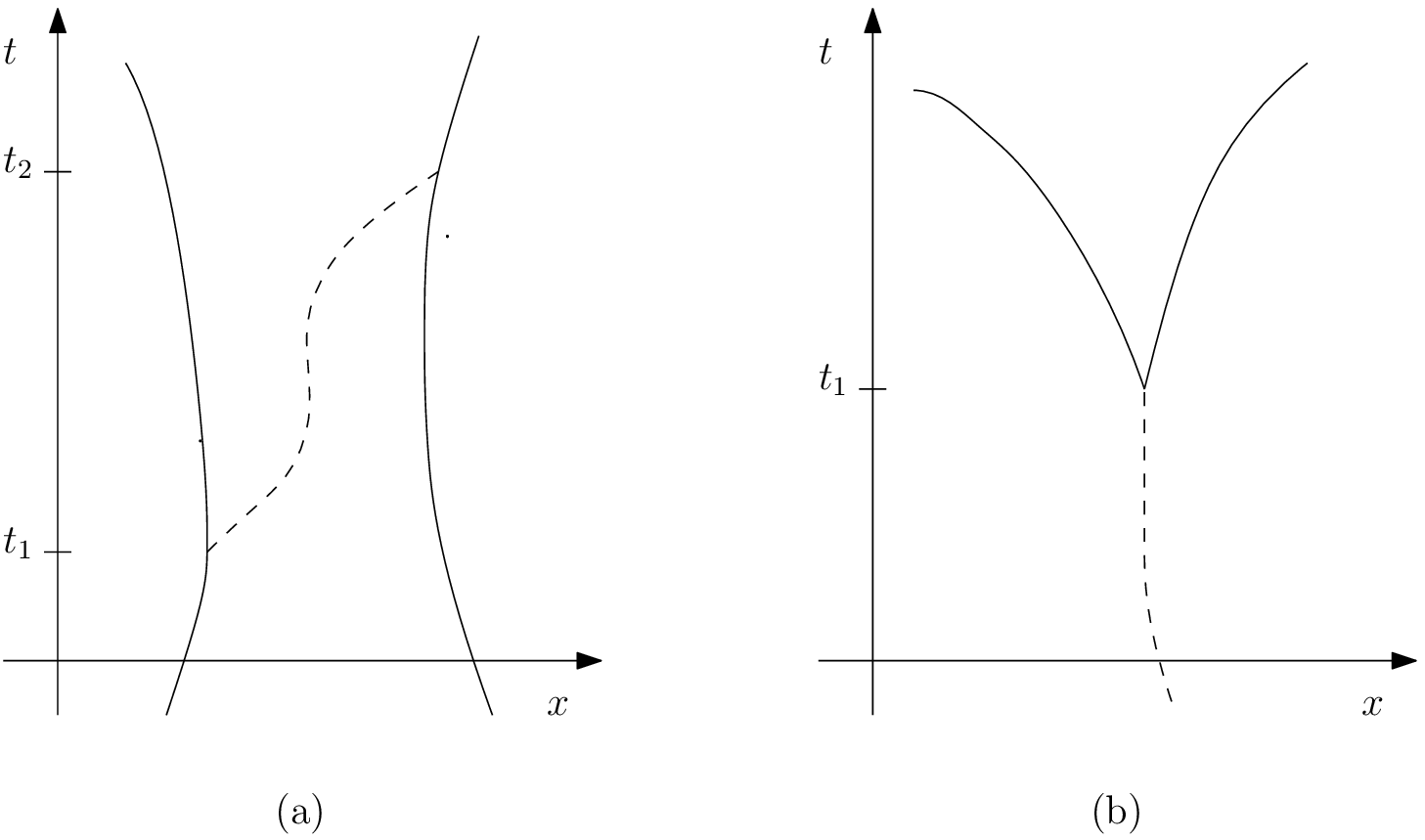}

Picture taken from \cite{Durr:2004aa}.
\end{center} 

Destinations and times of these jumps are the stochastic elements of BTQFT, and the transitions are described as Markov processes, since they do not depend on the past history of the particle configuration. Interestingly, this theory preserves Born's distribution, thus, it is empirically equivalent to the standard version of QFT (cf.\ \cite{Durr:2003aa, Durr2005} for technical details). 

BTQFT has remarkable features: it is ontologically well-defined and does not contain ill-defined notions within its axioms, as required by the primitive ontology programme. Moreover, every physical phenomenon is reduced and explained in terms of particles and their trajectories, as in BM. Nevertheless, BTQFT presents some metaphysical weaknesses which are absent in DQFT. Firstly, the ontological status of wave function, which should be considered a nomological entity, is not completely clear. The main argument to support the nomological view is given by the analogy with the classical Hamiltonian, a function in phase space which generates---via the well-known Hamilton equations---a vector field in such a space determining velocities and momenta of classical particles. According to \cite{Goldstein:2013} the wave function plays an analogous role for the motion of the Bohmian corpuscles. However, this analogy is not completely convincing. On the one hand,  as \cite{Romano:2020} (Section 2.1) underlines, the Hamiltonian is constructed \emph{from} definite properties of classical particles, as for instance their positions, velocities and masses; moreover the Hamiltonian should be considered as a useful mathematical tool which provides a simplified representation of the dynamics of classical particles---given that it is formally simpler to solve Hamilton equations with respect to a set of differential equations of Newton's mechanics. To this regard, Romano notes that the Hamiltonian is not indispensable in order to describe the dynamics of classical particles, since it can be given in Newtonian terms. On the other hand, the wave function in BM and BTQFT is not constructed starting directly from properties of Bohmian particles. Being the wave function a solution of the Schr\"odinger equation, it depends on the specific form of the Hamiltonian at hand, meaning that the positions and velocities of Bohmian corpuscles are relevant in order to define the Hamiltonian of the particles' configuration under consideration, but they are not useful to construct the wave function of such a configuration.
Moreover, the latter cannot be eliminated or dispensed from the formal structures of Bohmian mechanics or Bell-type QFT. Contrary to the Hamiltonian case, there are physical phenomena which would remain without explanation dispensing the wave function from such theories, as for instance the interference pattern in the double slit experiment (cf.\ \cite{Bell:2004aa}, p.\ 191). Nonetheless, it is not clear in the metaphysical framework proposed in \cite{Goldstein:2013} how a nomological entity not defined in spacetime---and which does not physically guide the motion of Bohmian particles as a proper pilot-wave would do (cf.\ \cite{Bohm:1952aa})---can have such physical effects.\footnote{It is worth noting that the ontology of Bohmian mechanics as presented by D\"urr, Goldstein and Zangh\`i in \cite{Durr:2013aa} differs remarkably from the causal approach contained in \cite{Bohm:1952aa}. Whereas in the former theory matter is only represented by particles in motion in space, in the latter theory the ontology is richer. Not only Bohm postulates a particle ontology, but he also proposed a realist view of the wave function, considering it a physical field in three-dimensional space acting directly on the particles, and derived the existence of a quantum potential in addition to the classical potential. Hence, the representation of the physical reality is metaphysically different in these two theories.} Thus, the argument by analogy in support of the nomological view of the wave function in BM and BTQFT would require stronger evidence. 

Another issue with the nomological view comes from cosmology. It is useful to recall that according to the nomological view, only the wave function of the entire universe is a law, and such an object evolves dynamically according to a static equation, i.e.\ the already mentioned Wheeler-DeWitt equation which reads $H|\Psi\rangle=0$. Here it is crucial to underline that, given the current knowledge of quantum theories of gravity, it is still not clear how to combine properly degrees of freedom related with the geometry of spacetime, and those related with material objects. Alternatively stated, there is a tension between the ontological content of Bohmian mechanics---which is the framework where the discussion about the nomological view takes place---and that of the Wheeler-DeWitt equation. More precisely, the universal wave function of BM is the wave function whose argument is the system of $N$ particles composing the entire universe, whereas the universal wave function appearing in the Wheeler-DeWitt equation differs remarkably from an ontological perspective, being a functional of all field configurations definable on spacetime. Moreover, the Hamiltonian acting on the wave functional in the Wheeler-DeWitt equation does not determine the dynamical evolution of physical systems as in non-relativistic quantum theory. Hence, (i) the physical content and metaphysical interpretation of these two universal wave functions is notably different, and (ii) given that the Hamiltonian in the Wheeler-DeWitt equation does not describe the motion of particles' configuration, the analogy between the universal wave function in BM and BTQFT and the Hamiltonian in classical mechanics cannot find sound justifications from cosmological considerations based on the Wheeler-DeWitt equation.

Furthermore, as \cite{Romano:2020}, p.\ 9 stresses, given the current cosmological knowledge (i) it is not clear what is the correct and realistic dynamics for our universe, therefore, it is a risky metaphysical move to rely on speculative features of cosmological models in order to defend a nomological view of $\Psi$, and (ii) even admitting that the universal wave function would turn out to be static, its form would depend on boundary conditions of a given cosmological model.\footnote{We thank Antonio Vassallo for this remark (personal communication).} However, one would expect that $\Psi$ as fundamental law of nature would be necessary and non-dependent on such boundary conditions---which are usually freely chosen---imposed to cosmological models. Hence, given the above arguments, the hypothesis claiming that the wave function of the universe is a nomological entity is highly speculative and would require stronger defence.

Secondly, subsystems of the universe are described by effective wave functions $\psi$, entities which Goldstein and Zangh\`i defined ``quasi-nomological'', however, the authors left unexplained the meaning of such notion, which still remains metaphysically obscure.\footnote{For a similar position cf.\ \cite{Belot:2012aa}, p.\ 75.} Especially in the context of BTQFT, where particles are randomly created and annihilated, it is not clear what happens to their effective guiding waves, since also these objects seems to be created and destroyed. In picture (a) above, the effective guiding wave function of the photon is created at time $t_1$---such function cannot be the same of the electrons since it must be defined in a  symmetric Fock space---and annihilated at time $t_2$ with the particle, nonetheless, a quasi-nomological entity, defined from the universal $\Psi_t$, should not come into existence or cease to exist, since a law must remain unconditioned by what happens to the Bohmian particles. 

In DQFT we have a clear ontological picture given in terms of particles which are not guided by pilot-waves; indeed, density matrices or unravelings of the QME do not physically determine the motion of the quantum objects in space in the precise sense that they do not physically and causally guide the motion of particles in spacetime as pilot-waves are suppose to do. As we already noted in Section 2.3, the dynamics of DQFT can be written either in terms of a QME or in terms of unravelings, providing a different representation for the particles' motion. Thus, this theory does not include entities with a metaphysically obscure status. In the second place, contrary to BTQFT, dissipative QFT provides an explanation for the stochasticity of its dynamics, since in this theoretical framework the irreversible behavior is originated by the interaction of the particles with a heat bath, whereas in BTQFT the stochastic jumps are fundamental, unexplained facts of nature.\footnote{DQFT's dynamics if written in terms of unravelings of the QME presents similarities w.r.t.\ the description of creation/annihilation events in BTQFT, since both theories introduce stochastic interruptions of a continuous evolution in Fock space. However, such jumps are due to remarkably different reasons as noted above.} Finally, DQFT provides new tools and instruments for numerical simulations, whereas such applicative aspect of the theory is missing in BTQFT.

\subsection{DQFT and the Dirac Sea Picture}

The Dirac sea picture has been introduced in the context of the pilot wave theory by \cite{Bohm:1953aa} to extend his causal approach to QFT\footnote{See \cite{Bohm:1993aa}, Chapter 12, for a more extended discussion of this approach; for a philosophical discussion cf.\ \cite{Deckert:2019aa}.}; successively, the DS hypothesis has been developed by several supporters of BM as for instance in \cite{Colin:2003aa, Colin:2003ab}, \cite{Colin:2007aa}, \cite{Deckert2010a}. Contrary to the case of BTQFT and DQFT, this is a deterministic theory which postulates an ontology of permanent particles, whose number remains constant in time, providing a different explanation to the phenomena of particle creation and annihilation. Moreover, according to the DS, only fermions exist, since it is the minimal ontological commitment able to explain measurement outcomes and more generally all the empirical data available, as claimed by \cite{Bell:1986aa}. 

To sketch the Dirac sea model we consider only the electron sector of the SM and electrodynamic interactions---interactions with other types of particles' sectors of the SM are modeled by an ``effective'' time-dependent external potential.\footnote{Considering only electrodynamic interactions among electrons, they would repel each other implying that the spatial extensions among them become larger and larger giving rise to an unphysical behavior. The external potential modeling the interaction between the rest of the particles and the electrons constrains the spatial extensions of the latter. For details see \cite{Deckert:2019aa}, Section~2.} Furthermore, the universe is assumed to have finite volume, and the electrons' momenta are restricted to be lower than some ultraviolet momentum cutoff $\Lambda$. The two last conditions have to be introduced to obtain a mathematically well-behaved model. To cast the DS picture in the Bohmian framework we have to specify the following dynamical laws; the wave function evolves according to the Schr\"odinger equation \eqref{SE}, but in this case the full Hamiltonian $H$ has the particular form:
\begin{equation}
\label{DSham}
H=\sum^N_{k=1}\Big{(}H^0_k(q_k)+V_k(t, q_k)+H^I_k(q_k)\Big{)},
\end{equation}
\noindent being constituted by the following terms: 
\begin{itemize}
   \item the free Hamiltonian $H^0_k(q)=1^{\otimes(k-1)}\otimes{H^0(q)}\otimes{1}^{\otimes(N-k)}$, where the $H^0(q)=-ic\alpha\cdot\nabla_q+\beta{m}c^2$. Here $\alpha, \beta$ are $4\times4$-matrices of Dirac's equation;
   \item the time-dependent potential $V_k(t,q_k)$ modeling the effective interaction of all the particles on the $k^{th}$ electron;
   \item the interaction Hamiltonian $H^I_k=\frac{1}{2}\sum_{j\neq{k}}U(q-q_j)$. The interaction among electrons is modeled by the Coulomb potential $U(q)=\frac{e^2}{4\pi\epsilon_0}|q|^{-1}$: $\epsilon_0$ is the electric constant, while $e$ represents the electron's charge.  
\end{itemize}

\noindent The particles follow continuous trajectories in space according to the guiding equation:
\begin{equation}
\label{DSvelocity}
v_t(Q)=c\Big{(}\frac{j^{(k)}_t(Q)}{\rho_t(Q)}\Big{)}_{k=1,\dots, N}
\end{equation}
\noindent{where} $Q$ is the actual configuration of $N$ electrons (with well-defined positions), $\rho(Q), j^{(k)}_t(Q)$ are respectively the probability density and the quantum current generated by $\Psi_t$, $c$ represents the speed of light. The DS picture is empirically adequate since its dynamical equations preserve the Born's distribution, reproducing the statistics of standard QFT. 

The DS model is an interesting generalization of BM to the realm of QFT maintaining a deterministic dynamics and a fixed number of particles. Moreover, the DS picture does not contain ill-defined notions, and explains physical phenomena in terms of the motion of Bohmian particles guided by the $\psi$ function as in BM. Nonetheless, such theory presents some conceptual problems. 

Firstly, the nature of the $\Psi$ function is metaphysically obscure as already mentioned in the case of BTQFT. Secondly, a consequence of the DS picture is that we have a surplus of non-interacting electrons occupying the sea of negative energy states. In this theory the vacuum is full of particles homogeneously distributed, specifically it is a sum of positive and negative energy states particles which naturally split the total Hilbert space into two subspaces $\mathcal{H}_-$ and $\mathcal{H}_+$, representing positive and negative energy particles respectively---it is worth stressing that the interpretation of the negative energy states is not problematic, since energy is defined only as a parameter useful to disambiguate the species to which a certain particle belongs to, and not as an inherent property of the Bohmian particles\footnote{For the lack of space we assume familiarity with the metaphysics of propertyless particles in BM (cf.\ \cite{Goldstein:2005b, Goldstein:2005a}, \cite{Esfeld:2015aa}). For a systematic introduction to the metaphysics of the DS see \cite{Esfeld:2017}.}, hence, the negative energies individuate the motion of the \emph{positrons}. The DS picture prescribes that all the negative energy states in the vacuum are occupied in virtue of the Pauli exclusion principle, so that positive energy particles do not fall into lower and lower energy states. Consequently, our world has a very high density of electrons in space, although such particles are in principle not observable. All these issues are absent in DQFT, since on the one hand, there is no pilot wave determining the motion of the particles---as already stated above---and on the other hand, there is no surplus of an invisible density of particles in physical space. Therefore, it is possible to  conclude that DQFT provides a simpler ontological description of the world w.r.t.\ the DS picture.

In the third place, this latter theory does not allow for a variable number of particles although the SM phenomenology suggests that the particle number is not constant. Indeed, the formalism of the DS approach defined on a $N$-particle Hilbert space does not contain particle creation and annihilation operators. However, one may recast this model introducing the Fock space formalism, allowing for the treatment of a variable number of particles. In the DS case, $\mathcal{F}$ would keep track of the wave function excitations with respect to the vacuum state. In virtue of the isomorphism between the ($N$-particle sector of the) Fock space and the $N$-particle Hilbert space representations, one may rewrite the dynamics generated by \eqref{DSham} in terms of the creation and annihilation operators, obtaining the canonical second-quantized Hamiltonian. According to DS picture, however, the terms ``creation'' and ``annihilation'' refer to a specific mathematical formalism, and not to physical events in space. What happens at the physical level is only that the particles arrange spatially in a certain way that can be described by either a Fock space formalism, which provides an efficient description of the variation of the vacuum excitations---and not the variation of the particle number---or by the $N$-particles Hilbert space, where the dynamics of every single individual object is specified.

Hence, the DS approach must provide a very articulate (and artificial) explanation of the particle creation and annihilation phenomena, whereas in DQFT such events are naturally explicated by arguments from nonequilibrium thermodynamics. 

In the fourth place, postulating an ontology with fixed number of particles may be disadvantageous given that it may have negative consequences for the empirical adequacy of this model. Indeed, taking seriously into account the predictions of the standard model, it is possible to note that for sufficiently high energies the latter prescribes the violation of the fermion number conservation. Such a prediction stands in clear contrast with the fundamental assumption of the DS model (cf.\ \cite{Colin:2007aa}, Sections 5 and 6). Hence, this fact may not only undermine the empirical adequacy of the DS proposal, but also it would show---contrary to the claims of its supporters---that it is not completely empirically equivalent to the standard model of particle physics. Such a problem, instead, is avoided by DQFT (and also by BTQFT)---in virtue of its ontology and its the stochastic dynamics---which allows for a variable number of particles.

Finally, the $H^I$ term in \eqref{DSham} contains explicit long-range interactions among point-particles, meaning that these objects can interact at a distance. This is an ontologically high price  price to be paid in order to keep perfect localization of the particles' positions in space. Alternatively stated, Bohmian mechanics in general, and the DS picture in particular, postulate an ontology of particles always well-localized in space that can interact non-locally at a distance; on the contrary, in DQFT one has exclusively local interactions (implemented by momentum conservation in collisions), although particles are not localized. Hence, we are confronted with the strange choice between ``localized particles and non-local interactions'', or ``non-localized particles but local interactions''.\footnote{Recall that in DQFT although particles are not strictly localized in space, it is assumed that they exist and are somewhere in spacetime, as stressed in Section~\ref{Discussion}.} Currently, the physicist community tends to agree with the second option, which is embedded in the structure of DQFT. Finally, the applicative aspect of DQFT is missing in the DS theory. 

\section{Conclusion}
\label{Conclusion}

In this essay we have introduced a new alternative formulation of QFT with a clear particle ontology and a stochastic dynamics based (i) on mathematically and physically rigorous notions from nonequilibrium thermodynamics, and (ii) on sound metaphysical assumptions, taking the best of both worlds, the algebraic and the Bohmian perspectives---although DQFT remarkably differs from these approaches. 

In the second place, with DQFT we have practically shown that also effective QFTs can provide robust and unambiguous ontological pictures of the objects and processes which are assumed to describe the physical world at certain energy/length scales, in agreement with the current philosophical literature on the foundations of QFT (cf.\ to this regard \cite{Williams:2017}, \cite{JFraser2018}, \cite{Egg2017}). Consequently, we also have shown that a realistic interpretation of the ontology of the standard model of particle physics is possible. Referring to this, after having introduced the particle ontology of DQFT, and explained how this theory is motivated by the experimental evidence available in high energy physics, we showed how such a theoretical framework avoid the ontological implications of Haag's theorem in a new way.

Furthermore, after having discussed the main implications and consequences of DQFT, we compared it with BTQFT and the DS picture---the most developed Bohmian QFTs with a particle ontology. 
We argued that, although there are similarities between these perspectives, the dissipative approach is not plagued by some important conceptual puzzles affecting these theories, since it neither contains a pilot wave physically guiding the motion of the quantum particles (whose metaphysical status is not completely clarified), nor its ontology entails any surplus of unobservable particles, but rather it provides more substantial arguments---taken from nonequilibrium thermodynamics---to explain the phenomena of particles creation and annihilation w.r.t.\ both BTQFT and the DS model and finally, it has an applicative aspect, providing new tools for numerical simulations, which is absent in both these frameworks. 

Hence, it is possible to conclude that there are sound motivations to consider such an approach as a possible, valid alternative to QFT; thus, for the reasons explained in this essay, we are convinced that DQFT may deserve attention from the community working on the philosophical foundations of the quantum theory of fields.

\clearpage
\bibliographystyle{apalike}
\bibliography{dissipative_QFT}

\begin{thebibliography}{}

\bibitem[Allori, 2013]{Allori:2013ab}
Allori, V. (2013).
\newblock Primitive ontology and the structures of fundamental physical
  theories.
\newblock In Albert, D.~Z. and Ney, A., editors, {\em The Wave Function: Essays
  on the Metaphysics of Quantum Mechanics}, chapter~2, pages 58 -- 75. Oxford
  University Press.

\bibitem[Auyang, 1995]{Auyang:1995}
Auyang, S.~Y. (1995).
\newblock {\em How is Quantum Field Theory Possible?}
\newblock Oxford University Press.

\bibitem[Baker, 2009]{Baker2009}
Baker, D.~J. (2009).
\newblock Against field interpretations of quantum field theory.
\newblock {\em British Journal for the Philosophy of Science}, 60:585--609.

\bibitem[Barrett, 2014]{Barrett:2014aa}
Barrett, J.~A. (2014).
\newblock Entanglement and disentanglement in relativistic quantum mechanics.
\newblock {\em Studies in History and Philosophy of Science Part B: Studies in
  History and Philosophy of Modern Physics}, 48:168--174.

\bibitem[Bell, 1986]{Bell:1986aa}
Bell, J.~S. (1986).
\newblock Beables for quantum field theory.
\newblock {\em Physics Reports}, 137:49--54.

\bibitem[Bell, 1987]{Bell:2004aa}
Bell, J.~S. (1987).
\newblock {\em Speakable and unspeakable in quantum mechanics}.
\newblock Cambridge: Cambridge University Press.

\bibitem[Belot, 2012]{Belot:2012aa}
Belot, G. (2012).
\newblock Quantum states for primitive ontologists: a case study.
\newblock {\em European Journal for Philosophy of Science}, 2(1):67--83.

\bibitem[Bigaj, 2018]{Bigaj:2018}
Bigaj, T. (2018).
\newblock Are field quanta real objects? {S}ome remarks on the ontology of
  quantum field theory.
\newblock {\em Studies in History and Philosophy of Science Part B: Studies in
  History and Philosophy of Modern Physics}, 62:145--157.

\bibitem[Bohm, 1952]{Bohm:1952aa}
Bohm, D. (1952).
\newblock {A suggested interpretation of the quantum theory in terms of
  ``hidden'' variables. I, II}.
\newblock {\em Physical Review}, 85(2):166 --193.

\bibitem[Bohm, 1953]{Bohm:1953aa}
Bohm, D. (1953).
\newblock {Comments on an article of Takabayasi concerning the formulation of
  quantum mechanics with classical pictures}.
\newblock {\em Progress of Theoretical Physics}, 9(3):273 -- 287.

\bibitem[Bohm and Hiley, 1993]{Bohm:1993aa}
Bohm, D. and Hiley, B. (1993).
\newblock {\em The Undivided Universe: An Ontological Interpretation of Quantum
  Theory}.
\newblock Routledge.

\bibitem[Born et~al., 1926]{Born:1926}
Born, M., Heisenberg, W., and Jordan, P. (1926).
\newblock {Zur Quantenmechanik II}.
\newblock {\em Zeitschrift f{\"u}r Physik}, 35(8-9):557--615.

\bibitem[Breuer and Petruccione, 2002]{Petruccione:2002}
Breuer, H.-P. and Petruccione, F. (2002).
\newblock {\em {The Theory of Open Quantum Systems}}.
\newblock Oxford University Press.

\bibitem[Butterfield and Bouatta, 2015]{Butterfield:2015}
Butterfield, J. and Bouatta, N. (2015).
\newblock {Renormalization for philosophers}.
\newblock In Bigaj, T. and W{\"u}thrich, C., editors, {\em {Metaphysics in
  Contemporary Physics}}, volume 104, pages 437--485. Pozna\'n Studies in the
  Philosophy of the Sciences and the Humanities.

\bibitem[Colin, 2003a]{Colin:2003aa}
Colin, S. (2003a).
\newblock {A deterministic Bell model}.
\newblock {\em Physics Letters A}, 317(5-6):349--358.

\bibitem[Colin, 2003b]{Colin:2003ab}
Colin, S. (2003b).
\newblock Beables for quantum electrodynamics.
\newblock {\em Annales de la Fondation Louis de Broglie, Manuscrit}, pages 1 --
  23.

\bibitem[Colin and Struyve, 2007]{Colin:2007aa}
Colin, S. and Struyve, W. (2007).
\newblock {A Dirac sea pilot-wave model for quantum field theory}.
\newblock {\em Journal of Physics A}, 40(26):7309--7341.

\bibitem[Deckert et~al., 2019]{Deckert:2019aa}
Deckert, D., Esfeld, M., and Oldofredi, A. (2019).
\newblock {A persistent particle ontology in terms of the Dirac sea}.
\newblock {\em British Journal for the Philosophy of Science}, 70(3):747--770.

\bibitem[Deckert et~al., 2010]{Deckert2010a}
Deckert, D.-A., D\"{u}rr, D., Merkl, F., and Schottenloher, M. (2010).
\newblock {Time-evolution of the external field problem in quantum
  electrodynamics}.
\newblock {\em {Journal of Mathematical Physics}}, {51}(12):122301.

\bibitem[Duane and Kogut, 1986]{Duane:1986}
Duane, S. and Kogut, J. (1986).
\newblock {The theory of hybrid stochastic algorithms}.
\newblock {\em Nuclear Physics B}, 275:398--420.

\bibitem[Duncan, 2012]{Duncan:2012aa}
Duncan, A. (2012).
\newblock {\em The Conceptual Framework of Quantum Field Theory}.
\newblock Oxford University Press.

\bibitem[D{\"u}rr et~al., 2003]{Durr:2003aa}
D{\"u}rr, D., Goldstein, S., Tumulka, R., and Zangh{\`\i}, N. (2003).
\newblock Trajectories and particle creation and annihilation in quantum field
  theory.
\newblock {\em Journal of Physics A: Mathematical and General}, 36:4143--4149.

\bibitem[D{\"u}rr et~al., 2004]{Durr:2004aa}
D{\"u}rr, D., Goldstein, S., Tumulka, R., and Zangh{\`\i}, N. (2004).
\newblock Bohmian mechanics and quantum field theory.
\newblock {\em Physical Review Letters}, 93:090402.

\bibitem[D{\"u}rr et~al., 2005]{Durr2005}
D{\"u}rr, D., Goldstein, S., Tumulka, R., and Zangh{\`\i}, N. (2005).
\newblock Bell-type quantum field theories.
\newblock {\em Journal of Physics A: Mathematical and General}, 38(4):R1--R43.

\bibitem[D{\"u}rr et~al., 2013]{Durr:2013aa}
D{\"u}rr, D., Goldstein, S., and Zangh{\`\i}, N. (2013).
\newblock {\em {Quantum Physics without Quantum Philosophy}}.
\newblock Springer.

\bibitem[Egg et~al., 2017]{Egg2017}
Egg, M., Lam, V., and Oldofredi, A. (2017).
\newblock {Particles, cutoffs and inequivalent representations: Fraser and
  Wallace on quantum field theory}.
\newblock {\em Foundations of Physics}, 47(3):453--466.

\bibitem[Esfeld and Deckert, 2017]{Esfeld:2017}
Esfeld, M. and Deckert, D.-A. (2017).
\newblock {\em {A minimalist Ontology of the Natural World}}.
\newblock Routledge.

\bibitem[Esfeld et~al., 2015]{Esfeld:2015aa}
Esfeld, M., Lazarovici, D., Lam, V., and Hubert, M. (2015).
\newblock The physics and metaphysics of primitive stuff.
\newblock {\em British Journal for the Philosophy of Science}, 68(1):133--162.

\bibitem[Falkenburg, 2007]{Falkenburg:2007aa}
Falkenburg, B. (2007).
\newblock {\em Particle Metaphysics. A Critical Account of Subatomic Reality}.
\newblock Springer.

\bibitem[Fraser, 2006]{Fraser2006}
Fraser, D. (2006).
\newblock Haag's theorem and its implications for the foundations of quantum
  field theory.
\newblock {\em Erkenntnis}, 64:305--344.

\bibitem[Fraser and Earman, 2008]{Fraser2008}
Fraser, D. and Earman, J. (2008).
\newblock The fate of ``particles' in quantum field theories with interactions.
\newblock {\em Studies in History and Philosophy of Science Part B: Studies in
  History and Philosophy of Modern Physics}, 38:841--859.

\bibitem[Fraser, 2018]{JFraser2018}
Fraser, J. (2018).
\newblock {\em Towards a realist view of quantum field theory}, chapter draft
  for the volume ``Scientific realism and the quantum''.
\newblock Oxford University Press, available at
  http://philsci-archive.pitt.edu/14716/.

\bibitem[Gardiner and Zoller, 2004]{Zoller:2004}
Gardiner, C. and Zoller, P. (2004).
\newblock {\em {Quantum Noise: A Handbook of Markovian and Non-Markovian
  Quantum Stochastic Methods with Applications to Quantum Optics}}.
\newblock Springer.

\bibitem[Goldstein et~al., 2005a]{Goldstein:2005b}
Goldstein, S., Taylor, J., Tumulka, R., and Zangh{\`\i}, N. (2005a).
\newblock Are all particles identical?
\newblock {\em Journal of Physics A: Mathematical and General},
  38(7):1567--1576.

\bibitem[Goldstein et~al., 2005b]{Goldstein:2005a}
Goldstein, S., Taylor, J., Tumulka, R., and Zangh{\`\i}, N. (2005b).
\newblock Are all particles real?
\newblock {\em Studies in History and Philosophy of Science Part B: Studies in
  History and Philosophy of Modern Physics}, 36(1):103--112.

\bibitem[Goldstein and Zangh{\`\i}, 2013]{Goldstein:2013}
Goldstein, S. and Zangh{\`\i}, N. (2013).
\newblock {Reality and the role of the wave function in quantum theory}.
\newblock In Albert, D.~Z. and Ney, A., editors, {\em {The Wave Function:
  Essays on the Metaphysics of Quantum Mechanics}}, chapter~4. Oxford
  University Press.

\bibitem[Gottlieb et~al., 1987]{Gottlieb:1987}
Gottlieb, S., Liu, W., Toussaint, D., Renken, R., and Sugar, R. (1987).
\newblock {Hybrid-molecular-dynamics algorithms for the numerical simulation of
  quantum chromodynamics}.
\newblock {\em Physical Review D}, 35:2531--2542.

\bibitem[Haag, 1955]{Haag:1955}
Haag, R. (1955).
\newblock On quantum field theories.
\newblock {\em Det Kongelige Danske Videnskabernes Selskab, Matematisk-fysiske
  Meddeleser}, 29(12):1--37.

\bibitem[Haag and Kastler, 1964]{Haag:1964}
Haag, R. and Kastler, D. (1964).
\newblock {An algebraic approach to quantum field theory}.
\newblock {\em {Journal of Mathematical Physics}}, 5:848--861.

\bibitem[Halvorson and Clifton, 2002]{Halvorson2002}
Halvorson, H. and Clifton, R.~K. (2002).
\newblock No place for particles in relativistic quantum theories?
\newblock {\em Philosophy of Science}, 69(1):1--28.

\bibitem[Hegerfeldt, 1998]{Hegerfeldt1998b}
Hegerfeldt, G. (1998).
\newblock {Instantaneous spreading and Einstein causality in quantum theory}.
\newblock {\em Annalen der Physik}, 7:716--725.

\bibitem[Kochen and Specker, 1967]{Kochen:1967}
Kochen, S. and Specker, E.~P. (1967).
\newblock {The problem of hidden variable in quantum mechanics}.
\newblock {\em Journal of Mathematics and Mechanics}, 17(1):59--87.

\bibitem[Loh et~al., 1990]{Loh:1990}
Loh, E., Gubernatis, J., Scalettar, R., White, S., Scalapino, D., and Sugar, R.
  (1990).
\newblock {Sign problem in the numerical simulation of many-electron systems}.
\newblock {\em Physical Review B}, 41(13):9301--9307.

\bibitem[MacKinnon, 2008]{MacKinnon2008}
MacKinnon, E. (2008).
\newblock The standard model as a philosophical challenge.
\newblock {\em Philosophy of Science}, 75:447--457.

\bibitem[Malament, 1996]{Malament1996}
Malament, D. (1996).
\newblock In defense of dogma: Why there cannot be a relativistic quantum
  mechanics of (localizable) particles.
\newblock In Clifton, R.~K., editor, {\em Perspectives on Quantum Reality},
  pages 1--11. Kluwer.

\bibitem[Osterwalder and Schrader, 1973]{Osterwalder:1973}
Osterwalder, K. and Schrader, R. (1973).
\newblock {Axioms for Euclidean Green's functions}.
\newblock {\em Communications in Mathematical Physics}, 31(2):83--112.

\bibitem[\"Ottinger, 2005]{Ottinger:2005}
\"Ottinger, H.~C. (2005).
\newblock {\em {Beyond Equilibrium Thermodynamics}}.
\newblock John Wiley \& Sons.

\bibitem[\"Ottinger, 2009]{Ottinger:2009}
\"Ottinger, H.~C. (2009).
\newblock {Dynamic renormalization in the framework of nonequilibrium
  thermodynamics}.
\newblock {\em Physical Review E}, 79:021124.

\bibitem[\"Ottinger, 2011]{Ottinger:2011}
\"Ottinger, H.~C. (2011).
\newblock {Dynamic coarse-graining approach to quantum field theory}.
\newblock {\em Physical Review D}, page 065007.

\bibitem[\"Ottinger, 2017]{Ottinger:2017}
\"Ottinger, H.~C. (2017).
\newblock {\em {A Philosophical Approach to Quantum Field Theory}}.
\newblock Cambridge University Press.

\bibitem[\"Ottinger, 2020a]{Ottinger:2020}
\"Ottinger, H.~C. (2020a).
\newblock {Composite Higer Derivative Theory of Gravity}.
\newblock {\em Physical Review Research}, 2:013190.

\bibitem[\"Ottinger, 2020b]{Ottinger:2020b}
\"Ottinger, H.~C. (2020b).
\newblock {Mathematical Structure and Physical Content of Composite}.
\newblock {\em Physical Review D}, 102:064024.

\bibitem[\"Ottinger et~al., 2020]{Ottinger:2020c}
\"Ottinger, H.~C., Peletier, M., and Montefusco, A. (2020).
\newblock {A Framework of Nonequilibrium Statistical Mechanics}.
\newblock {\em Journal of Non-Equilibrium Thermodynamics}, 45.

\bibitem[Pashby and \"Ottinger, 2020]{Pashby:2020}
Pashby, T. and \"Ottinger, H.~C. (2020).
\newblock {Quantum Jump Revival}.
\newblock {\em Unpublished Manuscript}.

\bibitem[Romano, 2020]{Romano:2020}
Romano, D. (2020).
\newblock {Multi-field and Bohm's theory}.
\newblock {\em Forthcoming in Synthese}, pages 1--30.

\bibitem[Schweber, 1994]{Schweber:1994}
Schweber, S. (1994).
\newblock {\em {QED and the Men Who Made It: Dyson, Feynman, Schwinger and
  Tomonaga}}.
\newblock Princeton University Press.

\bibitem[Struyve, 2010]{Struyve:2010aa}
Struyve, W. (2010).
\newblock Pilot-wave approaches to quantum field theory.
\newblock {\em Journal of Physics: Conference Series}, 306:012047.

\bibitem[Taj and \"Ottinger, 2015]{Ottinger:2015}
Taj, D. and \"Ottinger, H.~C. (2015).
\newblock {Natural approach to quantum dissipation}.
\newblock {\em Physical Review A}, 92:062128.

\bibitem[Teller, 1995]{Teller:1995aa}
Teller, P. (1995).
\newblock {\em An Interpretative Introduction to Quantum Field Theory}.
\newblock Princeton University Press.

\bibitem[Troyer and Wiese, 2005]{Troyer:2005}
Troyer, M. and Wiese, U. (2005).
\newblock {Computational complexity and fundamental limitations to fermionic
  quantum Monte Carlo simulations}.
\newblock {\em Physical Review Letters}, 94(17):170201.

\bibitem[Wallace, 2006]{Wallace2006}
Wallace, D. (2006).
\newblock In defence of naivet\'e: The conceptual status of {Lagrangian}
  quantum field theory.
\newblock {\em Synthese}, 151:33--80.

\bibitem[Wallace, 2011]{Wallace2011}
Wallace, D. (2011).
\newblock Taking particle physics seriously: A critique of the algebraic
  approach to quantum field theory.
\newblock {\em Studies in History and Philosophy of Science Part B: Studies in
  History and Philosophy of Modern Physics}, 42:116--125.

\bibitem[Wightman and G{\aa}rding, 1964]{Wightman:1964}
Wightman, A. and G{\aa}rding, L. (1964).
\newblock {Fields as Operator-valued Distributions in Relativistic Quantum
  Theory}.
\newblock {\em Arkiv f\"or fysik}, 28(129-189).

\bibitem[Williams, 2019]{Williams:2017}
Williams, P. (2019).
\newblock {Scientific realism made effective}.
\newblock {\em The British Journal for the Philosophy of Science},
  70(1):209--237.

\bibitem[Wilson, 1974]{Wilson:1974aa}
Wilson, K. (1974).
\newblock {Confinement of quark}.
\newblock {\em Physical Review D}, 10:2445--2459.

\bibitem[Wilson and Kogut, 1974]{Wilson:1974}
Wilson, K. and Kogut, J. (1974).
\newblock {The renormalization group and the $\epsilon$ expansion}.
\newblock {\em Physics Reports}, 12:75--200.

\end{thebibliography}
\end{document}